\RequirePackage{lineno}
\documentclass[%
 reprint,
superscriptaddress,
 amsmath,amssymb,
 aps,
]{revtex4-2}

\usepackage{graphicx}   
\usepackage{placeins}
\usepackage{xcolor}
\usepackage{natbib}
\usepackage{hyperref}

\usepackage{amsmath,amssymb,amsfonts}
\usepackage{xspace}
\usepackage{subfigure}

\usepackage[bottom]{footmisc} 

\begin{document}


\title{Constraining the CME in AVFD-simulated heavy-ion collisions using the Sliding Dumbbell Method}

\author{Jagbir Singh}
\affiliation{Instituto de Alta Investigaci\'on, Universidad de Tarapac\'a, Casilla 7D, Arica, 1000000, Chile}%

\author{Anjali Sharma}
\affiliation{Department of Physical sciences, Bose Institute, Kolkata 700091, India}%

\author{Ankita Nain}
\affiliation{Department of Physics, D.A.V. College, Sector 10, Chandigarh 160011, India}%

\author{Madan M. Aggarwal}
\email{aggarwal@pu.ac.in}
\affiliation{Department of Physics, Panjab University, Chandigarh 160014, India}%

\date{\today}

\begin{abstract}
The Anomalous Viscous Fluid Dynamics (AVFD) framework is utilized to generate  $^{197}_{79}Au+^{197}_{79}Au$, $^{96}_{44}Ru+^{96}_{44}Ru$, and $^{96}_{40}Zr+^{96}_{40}Zr$ collision events at $\sqrt{s_{\mathrm{NN}}}$ = 200 GeV to investigate the Chiral Magnetic Effect (CME). The CME signal is modulated through the axial charge per entropy density ($n_5/s$) in each event to produce data sets with varying CME signal strengths. Additionally, a 33$\%$ local charge conservation (LCC) is implemented in each event. These data sets are analyzed using CME-sensitive two- and three-particle correlators. Furthermore, the Sliding Dumbbell Method (SDM) is employed to identify potential CME-like events within each data set. The identified events selected using the SDM exhibit characteristics consistent with CME. The CME fraction in these events is quantified while accounting for background contributions.
\end{abstract}


\maketitle

\section{Introduction}
\label{sec:intro}
Quantum chromodynamics (QCD) predicts that meta-stable domains with fluctuating topological charges can induce changes in the chirality of quarks, leading to local CP violation under conditions of extremely high temperatures and/or densities, such as those prevailing during quark-gluon plasma (QGP) formation~\cite{1,2,3,4,Fukushima_2008}. In non-central heavy-ion collisions, the intense magnetic field generated by highly energetic spectator protons causes separation of oppositely charged particles along the system's angular momentum direction. This phenomenon is known as the Chiral Magnetic Effect (CME). The search for conclusive experimental evidence of the CME is one of the primary goals of the heavy-ion physics programs at both Relativistic Heavy Ion Collider (RHIC) and Large Hadron Collider (LHC). Such a discovery will have profound implications beyond heavy-ion physics, potentially marking a significant milestone in the field of physics as a whole. Consequently, extensive theoretical~\cite{Voloshin_2004,Bzdak_2013,Kharzeev_2016,Koch_2017,schenke_19,Zhao_2019,XU2020135706} and experimental~\cite{Li_20,PRL_103_251601,PhysRevC_81_054908,STAR_PhysRevC88,Adamczyk_2014,PhysRevLett_110_012301,Sirunyan_2018,Acharya_2018,JHEP_1029_2020,Abdallah_1079_2022,STAR_PRC105_2022,ALICE_PLB856,STAR_PLB839_2023,STAR_PRR6_2024} efforts have been devoted to probe the existence and properties of the CME. Number of methods~\cite{Voloshin_2004,Magdy_PRC97,Voloshin_PRC98_2018,Tang_CPC44,Aggarwal_SDM_2022} have been proposed to detect the CME signal in heavy-ion collisions. Efforts have also been made to estimate the CME signal using event shape engineering, pair invariant mass, etc.~\cite{Schukraft_PLB719_2013,Zhao_EPJC79_2019,Xu_PLB848_2024,Feng_PLB820_2021,Wu_2022fwz}.\par
The Anomalous Viscous Fluid Dynamics (AVFD) framework~\cite{Shi:2017cpu,Jiang:2016wve,Shi:2019wzi}, built upon the VISHNU bulk hydrodynamic evolution, is introduced to simulate the evolution of chiral fermion currents in the Quark-Gluon Plasma (QGP) during heavy-ion collisions.  The underlying evolution of the bulk medium is described by VISH2+1 hydrodynamics~\cite{Heinz:2015arc}, which provides a comprehensive view of the viscous behavior of the medium. 
The AVFD model integrates standard viscous hydrodynamics with anomalous fluid dynamics in a unified framework, taking into account crucial parameters such as initial conditions, magnetic fields, and viscous transport coefficients. This integration allows for a dynamic interplay between the evolution of the axial charge current and the bulk medium.\par 
The most widely used observable in the CME search is the “$\gamma$-correlator,” originally proposed by Voloshin~\cite{Voloshin_2004},
\begin{align}
  \gamma_{a, b} &= \langle cos(\phi_{a}+\phi_{b}-2\Psi_{RP})\rangle \notag \\
                &= \langle cos(\Delta\phi_{a})cos(\Delta\phi_{b}) \rangle - \langle sin(\Delta\phi_{a})sin(\Delta\phi_{b}) \rangle
  \label{eq:gamma}
  \end{align}
where, $\phi_{a}$ and $\phi_{b}$ are azimuthal angles of particles a and b, respectively, and $\Psi_{RP}$ represents the reaction plane angle. $\Delta\phi_{a}$ and $\Delta\phi_{b}$ represent azimuthal angles measured with respect to the reaction plane. Here the averaging $\langle$· · · $\rangle$ is performed over the pairs of particles and over events. Three-particle $\gamma$-correlator which is equivalent to the above $\gamma_{a, b}$ and is defined as~\cite{Voloshin_2004}:
\begin{equation}
  \gamma = \frac {\langle cos(\phi_{a}+\phi_{b}-2\phi_{c})\rangle}{v_{2, c}}
\end{equation}
where $\phi_{a}$ and $\phi_{b}$, and $\phi_{c}$ represent azimuthal angles of particles ``a'', ``b", and ``c", respectively. Here, a single particle ``c'' is used to measure the reaction plane angle and $v_{2, c}$ is the elliptic flow of particle c. In order to eliminate charge-independent correlation background mainly from global momentum conservation, the difference between the opposite-sign (OS) and same-sign (SS) charge pairs $\gamma$-correlators is considered,
\begin{equation}
  \Delta\gamma = \gamma_{OS}-\gamma_{SS}
\end{equation}
The $\Delta\gamma$ is sensitive to the preferential emission of positively and negatively charged particles to the opposite sides of the reaction plane.
\par
The reaction plane independent 2-particle $\delta$-correlator is also used which is given as:
\begin{align}
  \label{eq:delta1}
  \delta &= \langle cos(\phi_{a}-\phi_{b}) \rangle  \notag \\    
         &= \langle cos(\Delta\phi_{a})cos(\Delta\phi_{b})   \rangle + \langle 
               sin(\Delta\phi_{a})sin(\Delta\phi_{b}) \rangle
\end{align} 
From equations~\ref{eq:gamma} and~\ref{eq:delta1}, one can determine in-plane ($\langle cos(\Delta\phi_{a})cos(\Delta\phi_{b}) \rangle$) and out-of-plane ($\langle sin(\Delta\phi_{a})sin(\Delta\phi_{b}) \rangle$) correlations to check the preferential emission of charged particles. Other methods to search for the CME signal are viz., R observable~\cite{Magdy_PRC97}, participant and spectator planes method~\cite{Voloshin_PRC98_2018}, signed balance function~\cite{Tang_CPC44}, and sliding dumbbell method~\cite{Aggarwal_SDM_2022}.\par
In this analysis, the Sliding Dumbbell Method (SDM)~\cite{Aggarwal_SDM_2022} is employed to identify potential CME-like events. These events are further examined using the $\gamma$ and $\delta$ correlators to confirm that they display the expected characteristics of CME events. Additionally, background contributions are meticulously addressed through the use of charge-shuffle and correlated backgrounds. The structure of this paper is as follows: section~\ref{sec:SDM} provides a brief overview of the SDM, followed by a discussion on background estimation in section~\ref{sec:Bkg}. Section~\ref{sec:DataAna} presents various data samples utilized in this analysis, while results and discussion are given in section~\ref{sec:Results}. Finally, a summary is provided in section~\ref{sec:summary}.
\section{Sliding Dumbbell Method}
\label{sec:SDM}
The Chiral Magnetic Effect (CME) manifests as a separation of electric charge along the system's angular momentum axis, with positively charged particles moving in one direction and negatively charged particles in the opposite. This phenomenon motivates the search for back-to-back charge separation between positive and negative charges, with an overall charge excess asymmetry close to zero in the azimuthal plane. To identify potential CME-like events—those exhibit significant back-to-back charge separation, in the azimuthal plane, on an event-by-event basis in heavy-ion collisions, we developed the Sliding Dumbbell Method (SDM)~\cite{Aggarwal_SDM_2022}. This method is conceptually similar to the sliding window method used by the WA98 collaboration~\cite{WA98:2011qmq} to search for the disoriented chiral condensates.
\begin{figure}[htbp]
\centering
\includegraphics[width=.35\textwidth]{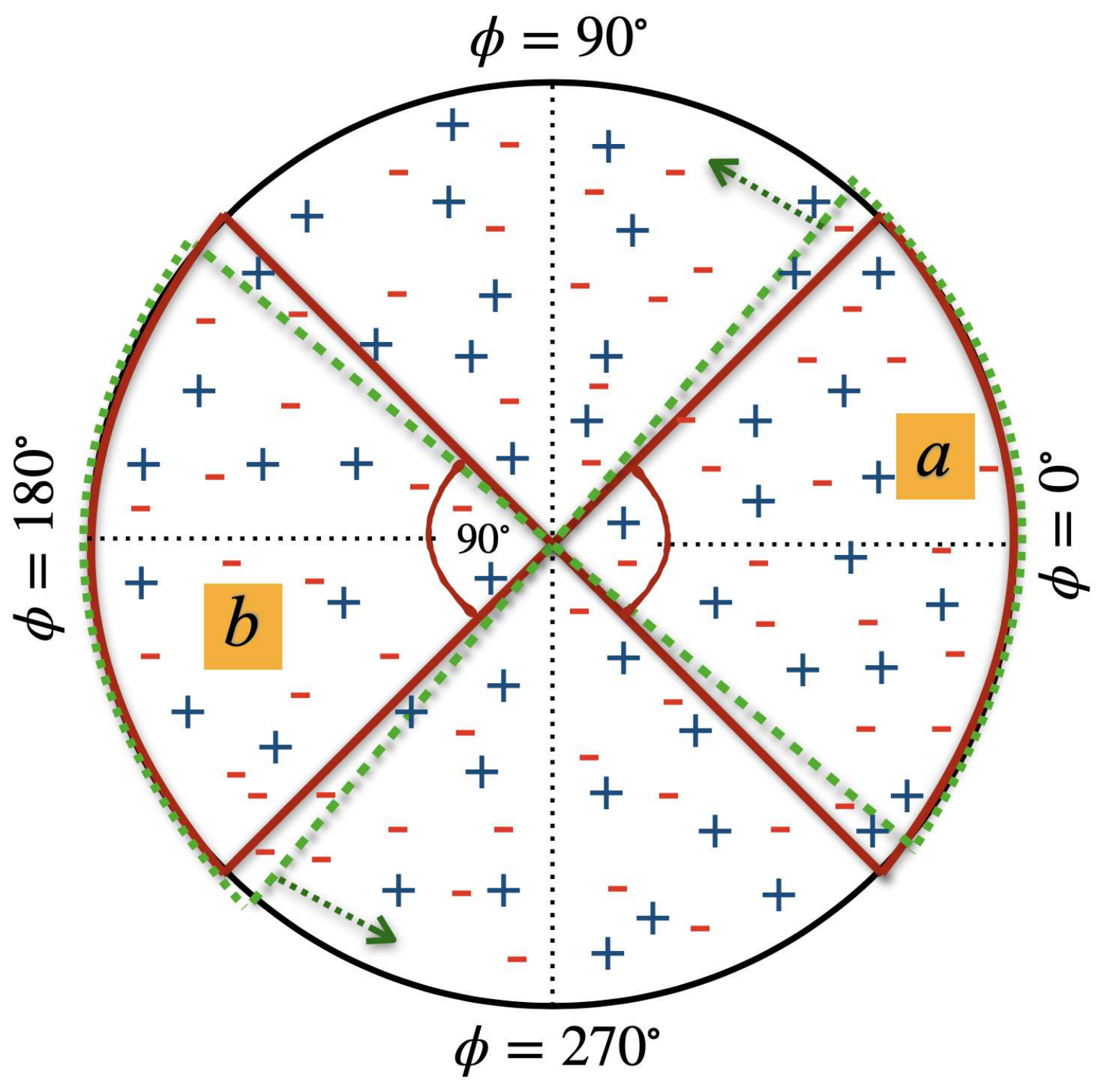}
\caption{Pictorial representation of the transverse plane with hits of positive (+) and negative (-) charge particles in an event. The dumbbell is shown in solid red line while the slid dumbbell is displayed in dotted green line.}
\label{fig:dumbbell}
\end{figure}
In the SDM, the azimuthal plane of each event is scanned by sliding a dumbbell-shaped region ($\Delta\phi=90{^\circ}$) in steps of $\delta\phi$ = $1^{\circ}$ as shown in Fig.~\ref{fig:dumbbell}. This approach allows for the identification of the region with the maximum back-to-back charge separation. To quantify this separation, we calculate $Db_{+-}$, which is the sum of the positive charge fraction on one side ``a'' ($f_{a}^{+}$) of the dumbbell and the negative charge fraction on the other side ``b'' ($f_{b}^{-}$) of the dumbbell, for each setting of the dumbbell across the azimuthal plane, i.e.,
  \begin{align}
    Db_{+-}  &= f_{a}^{+} +  f_{b}^{-}  \notag \\
            &= \frac{n_a^+}{(n_a^+ + n_a^-)} + \frac{n_b^-}{(n_b^+ + n_b^-)}
  \label{eq:db+-}
  \end{align}
  where, $n_{a}^{+}$($n_{b}^{+}$) and $n_{a}^{-}$($n_{b}^{-}$) represent the number of positive and negative charged particles, respectively, on side ``a'' (``b'') of the dumbbell. Both $f_{a}^{+}$ and $f_{b}^{-}$ are expected to be around $0.5$ for randomly emitted positively and negatively charged particles (i.e., $Db_{+-}\approx 1$). In case of an ideal Chiral Magnetic Effect (CME), however, all positively charged particles would move toward one side of the dumbbell ($f_{a}^{+}=1$), and all negatively charged particles toward the opposite side ($f_{b}^{-}=1$), resulting in $Db_{+-} = 2$. In practice, $f_{a}^{+}$ and $f_{b}^{-}$ can range between $0.5$ and $1.0$, leading to CME-like events with  $Db_{+-}$ values between $1$ and $2$. The fractional charge separation across the dumbbell ($f_{DbCS}$) can be defined as:
\begin{equation}
  f_{DbCS}  = Db_{+-} - 1 
  \label{eq:DbCS}
\end{equation}
Hereafter, $f_{DbCS}$ is referred as charge separation. Thus $Db_{+-}=2$ corresponds to $f_{DbCS}=1$, indicating  100$\%$ back-to-back charge separation, while $Db_{+-}=1$ (i.e., $f_{DbCS}=0$) indicates no back-to-back charge separation. Additionally, the charge excess asymmetry across the dumbbell, $Db_{+-}^{asy}$, is defined as:
\begin{equation}
  Db_{+-}^{asy}  = \frac{(n_{a}^{+}-n_{a}^{-})-(n_{b}^{-}-n_{b}^{+})}{(n_{a}^{+}-n_{a}^{-})+(n_{b}^{-}-n_{b}^{+})}
  \label{eq:dbasy}
\end{equation}
\newline 
Here, $n_{a}^{+}-n_{a}^{-}$ represents the positive charge excess on the ``a'' side of the dumbbell whereas ($n_{b}^{-}-n_{b}^{+}$) denotes the negative charge excess on the ``b'' side.  A value of  $Db_{+-}^{asy}=0$ indicates perfect charge separation between the two sides. In contrast,  $Db_{+-}^{asy}= \pm 1$ reflects a strong one-sided charge excess, with one side having nearly random charge distribution (e.g., $f_{a}^{+}\approx0.5$ or $f_{b}^{-}\approx0.5$). $Db_{+-}^{asy}$ can range from $-1$ to $1$. As its magnitude increases, one side of the dumbbell becomes increasingly dominated by either positively or negatively charged particles.

For both $Db_{+-}$ and $Db_{+-}^{asy}$, 360 values are obtained by sliding the dumbbell in steps of $\delta\phi=1^{\circ}$ across the azimuthal plane. The maximum value of $Db_{+-}$, termed $Db_{+-}^{max}$, is selected from those 360 values under the condition that $\mid Db_{+-}^{asy}\mid < 0.25$.
To ensure an approximately symmetric charge separation across the dumbbell, we apply a selection criterion of $\mid Db_{+-}^{asy}\mid < 0.25$.
Four distinct scenarios are considered, each involving 100 particles on either side of the dumbbell, but with varying distributions of positively and negatively charged particles. These are summarized in the Table I, along with the corresponding values of charge fractions $f_{a}^{+}$, $f_{b}^{-}$, $Db_{+-}$, $f_{DbCS}$, and $Db_{+-}^{asy}$.
\begin{itemize}
    \item Case I represents a typical event with random charge distribution, where $n_{a}^{+} = n_{a}^{-} = n_{b}^{-} = n_{b}^{+}=50$. Here, both $Db_{+-}^{asy}=0$ and $f_{DbCS}=0$, indicating no charge separation across the dumbbell.

    \item Case II illustrates an ideal CME-like event with complete charge separation, where $n_{a}^{+} = n_{b}^{-} =100$ and $n_{a}^{-} = n_{b}^{+} = 0$. This results in $Db_{+-}^{asy}=0$ and $f_{DbCS}=1$.

    \item Case III shows a scenario with approximately symmetric charge separation across the dumbbell, yielding $Db_{+-}^{asy}=0.05$ and $f_{DbCS}=0.53$.

    \item Case IV presents a highly asymmetric case with a strong positive charge concentration on one side ($f_{a}^{+}=0.90$) and a moderate negative charge fraction on the other ($f_{b}^{-}=0.55$). This leads to $f_{DbCS}=0.45$ and a large asymmetry $Db_{+-}^{asy}=0.77$. Such events are excluded from analysis, as only those with $Db_{+-}^{asy}<0.25$ are considered.
\end{itemize}
Cases II and III correspond to CME-like events. Case I, with $Db_{+-}=1$ and $f_{DbCS}=0$, indicates the absence of charge separation despite $Db^{asy}_{+-}=0$. In contrast, Case IV, characterized by $f^a_+=0.90$, reflects a charge excess on one side and is therefore not considered a CME-like event.

\begin{table}
  \caption{Lists four distinct scenarios, each involving 100 particles on either side of the dumbbell, but with varying distributions of positively and negatively charged particles. It also lists the values of $f_{a}^{+}$, $f_{b}^{-}$, $Db_{+-}$, $f_{DbCS}$, and $Db_{+-}^{asy}$.}
  \begin{center}
    \begin{tabular}{|c||c|c|c|c|c|c|c|c|}
      \hline
     & $n_{a}^{+}$ ($n_{a}^{-}$) & $n_{b}^{-}$ ($n_{b}^{+}$) & $f_{+}^{a}$ & $f_{-}^{b}$ & $Db_{+-}$ & $f_{DbCS}$ & $Db_{+-}^{asy}$ \\
    \hline
    \hline
      Case-I & 50 (50)  & 50 (50)  & 0.5 & 0.5 & 1 & 0 & 0\\
      \hline
      Case-II & 100 (0) & 100 (0)  & 1   & 1   & 2 & 1 & 0\\
      \hline
      Case-III & 75 (25)  & 78 (22) & 0.75  & 0.78  & 1.53 & 0.53 & -0.05\\
      \hline
      Case-IV  & 90 (10)  & 55 (45) & 0.90  & 0.55  & 1.45 & 0.45 & 0.77\\  
      \hline
    \end{tabular}
  \end{center}
  \label{tab:dbasy}
\end{table}

\begin{figure*}
  \centering
  \includegraphics[width=.43\textwidth]{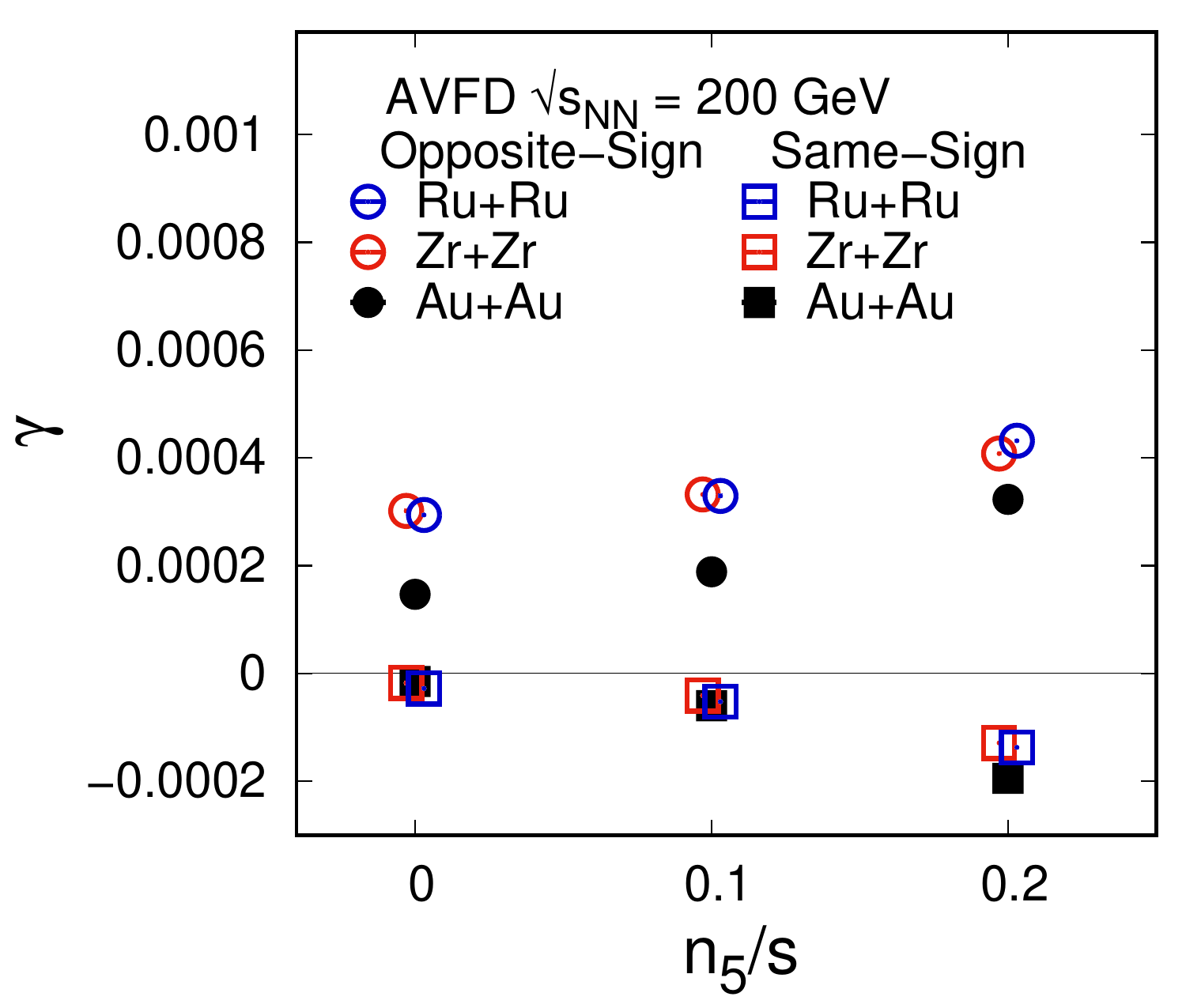}
  \includegraphics[width=.43\textwidth]{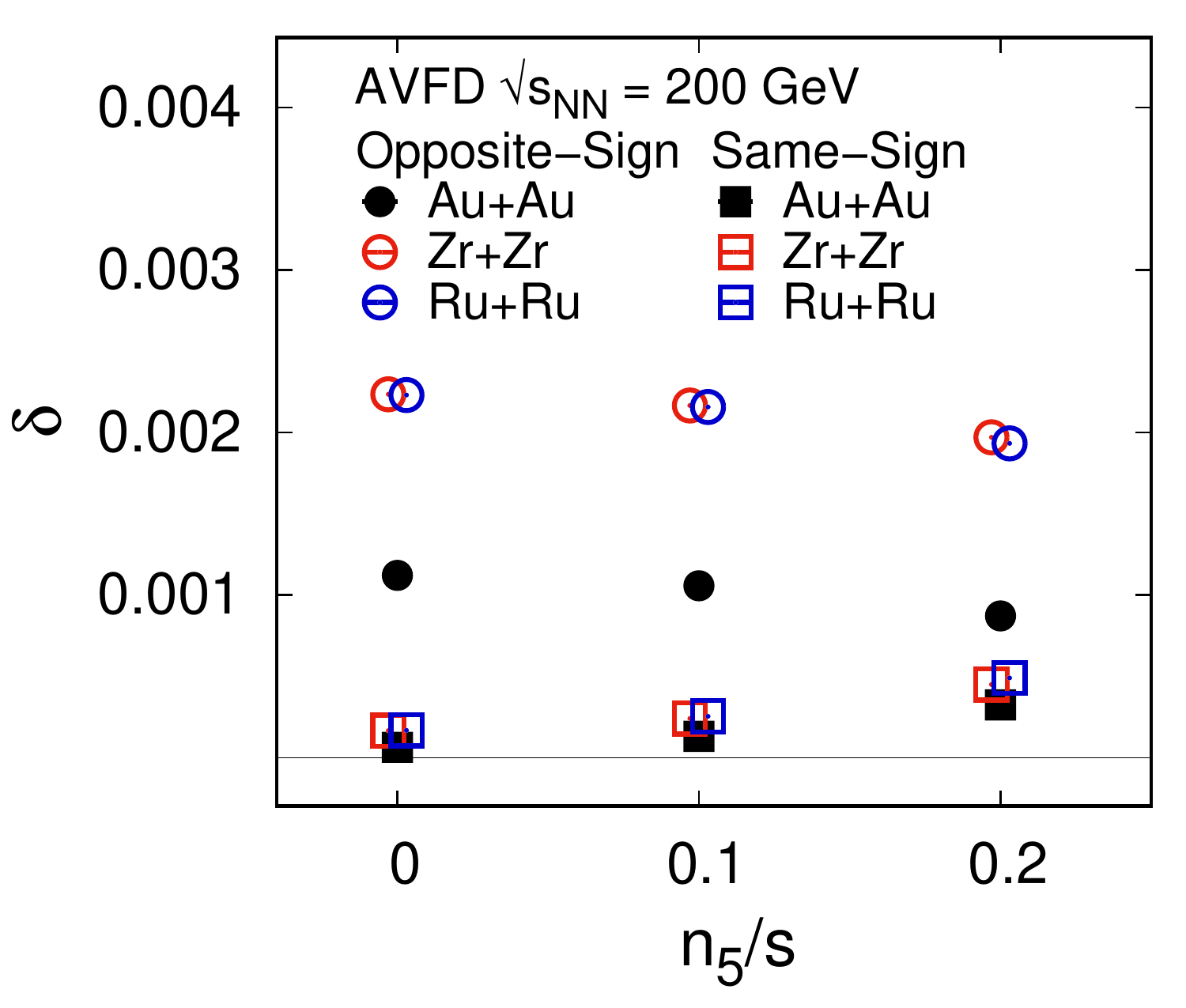}

  \includegraphics[width=.43\textwidth]{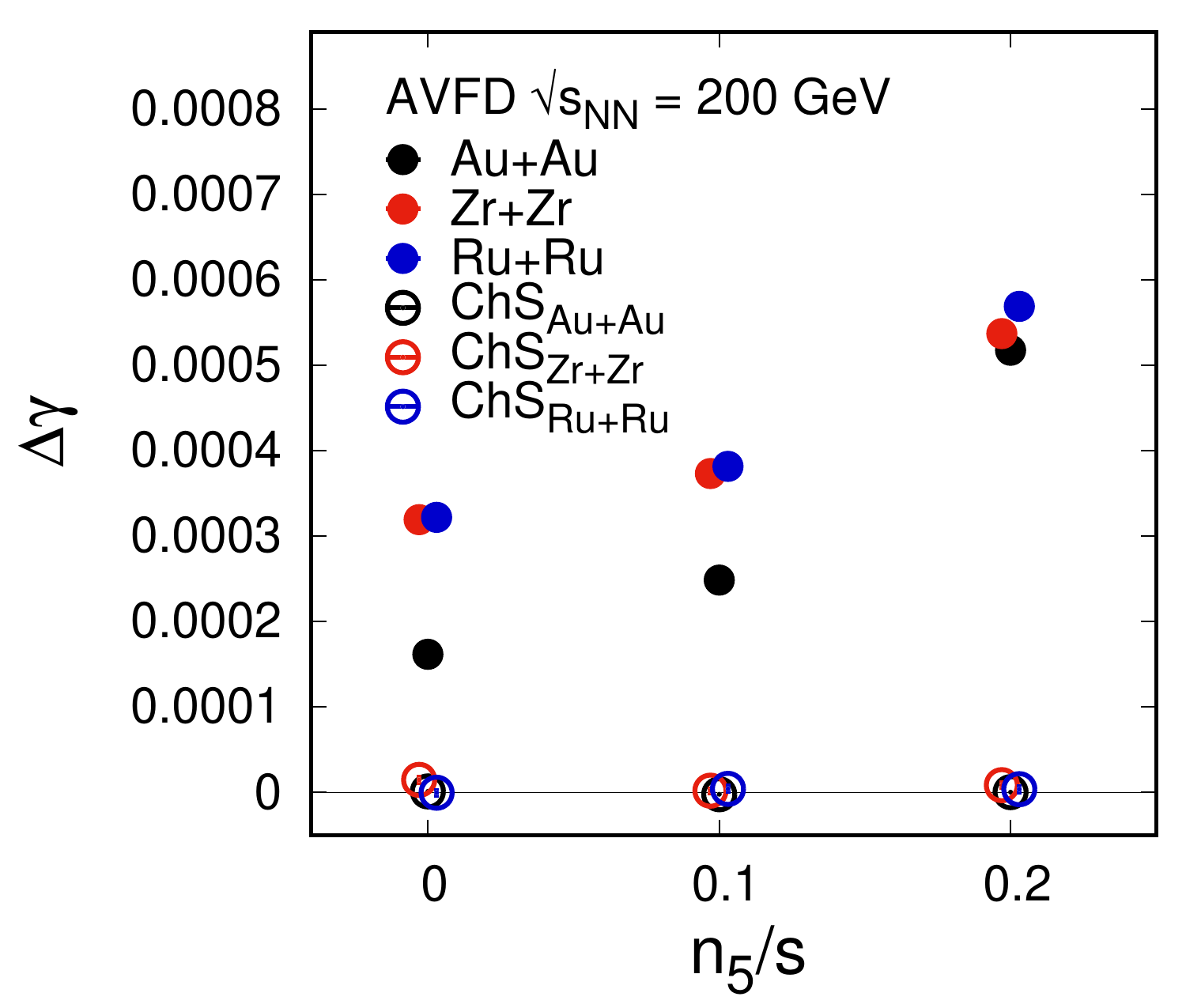}
  \caption{(Color Online) Three-particle $\gamma$-correlator (top left), two-particle $\delta$-correlator (top right), and $\Delta\gamma$ (bottom) for AVFD generated $Ru+Ru$, $Zr+Zr$ and $Au+Au$ collisions at $\sqrt{s_{\mathrm NN}}$ = 200 GeV versus $n_5/s$ for 30-40$\%$ collision centrality. The $\Delta\gamma$ plot (bottom) also includes charge shuffle ($\Delta\gamma_{ChS}$) values. Markers are slightly shifted along the x-axis for clarity. Statistical uncertainties are small and are within the marker size.}
  \label{fig:gamma}
\end{figure*}
\subsection{Background Estimation}
\label{sec:Bkg}
To calculate background contributions to the $\gamma$-correlator in different $f_{DbCS}$ percentile bins (will be discussed in sec.~\ref{sec:usingSDM}) using the SDM, we account for instances of higher charge separation occurring purely by chance while preserving the intrinsic correlations among particles. This is done by randomly shuffling the charges of particles in each event, keeping their momenta (i.e., $\theta$ and $\phi$) unchanged. The charge-shuffle sample for a given centrality is then analyzed in the same way as the original data set~\cite{Aggarwal_SDM_2022}. The $\gamma$ value for the charge-shuffle sample in a specific $f_{DbCS}$ bin is referred to as $\gamma_{ChS}$. Meanwhile, the charge correlations those were disrupted by the charge shuffling are recovered from the original events corresponding to a given $f_{DbCS}$ bins and termed as $\gamma_{Corr}$. Therefore, the total background contribution to the $\gamma$-correlator is expressed as:
\begin{equation}
  \gamma_{Bkg} = \gamma_{ChS} +\gamma_{Corr}
  \label{eq:bkg}
\end{equation}
This approach helps in estimating the background contribution to the $\gamma$-correlator.\par
The AVFD sample with $n_5/s=0$, including 33$\%$ LCC, can also serve as a background for the other samples with $n_5/s=0.1$ and $0.2$.

\section{Data Analyzed}
\label{sec:DataAna}
\begin{table}
  \caption{Lists number of events analyzed for AVFD generated $Ru+Ru$, $Zr+Zr$, and $Au+Au$ collisions at $\sqrt{s_{\mathrm{NN}}}$ = 200 GeV for different CME injections, for 30–40$\%$ collision centrality~\cite{avfd_star}.}
  \begin{center}
    \begin{tabular}{|c|c||c|c|c|}
      \hline
     \multicolumn{2}{|c||}{\textbf{AVFD}} & \multicolumn{3}{c|}{\textbf{Number of Events}} \\
    \hline
    \textbf{$n_5/s$} & \textbf{LCC} & \textbf{Au+Au} & \textbf{Ru+Ru} & \textbf{Zr+Zr} \\
    \hline
    \hline
      0.0 & 33$\%$ & $\sim$95 M & $\sim$58 M & $\sim$48 M\\
          \hline
      0.1 & 33$\%$ & $\sim$58 M & $\sim$49 M & $\sim$71 M\\
          \hline
      0.2 & 33$\%$ & $\sim$77 M & $\sim$50 M & $\sim$56 M\\
      \hline
    \end{tabular}
  \end{center}
  \label{tab:avfd}
\end{table}

In the simulations, the EBE-AVFD Beta1.0 version of the model is utilized~\cite{Feng_PLB820_2021,Shi:2017cpu}.
The AVFD framework modulates the CME signal through the axial charge per entropy density ($n_5/s$), which reflects the imbalance between right-handed and left-handed fermions introduced during the initial stage of each event. Another critical parameter in the model is the percentage of local charge conservation (LCC) within an event, which influences the background by dictating the proportion of positively and negatively charged partners emitted from the same fluid element relative to the total event multiplicity.\par
The axial charge per entropy density, $n_5/s=0.0$, 0.1 and 0.2, is used as input in the simulations for 30-40$\%$ collision centrality, as detailled in Ref.~\cite{avfd_star}. The AVFD generated $Au+Au$, $Ru+Ru$, and $Zr+Zr$ collisions at $\sqrt{s_{\mathrm NN}}~=~200$ GeV with 33$\%$ LCC in each event are analyzed, as listed in Table~\ref{tab:avfd}. Here, the samples with $n_5/s=0$ corresponds to no CME injection and includes only 33$\%$ LCC. This sample represents background due to local charge conservation without any CME signal.
\begin{figure*}[htbp]
  \centering
  \includegraphics[width=.44\textwidth]{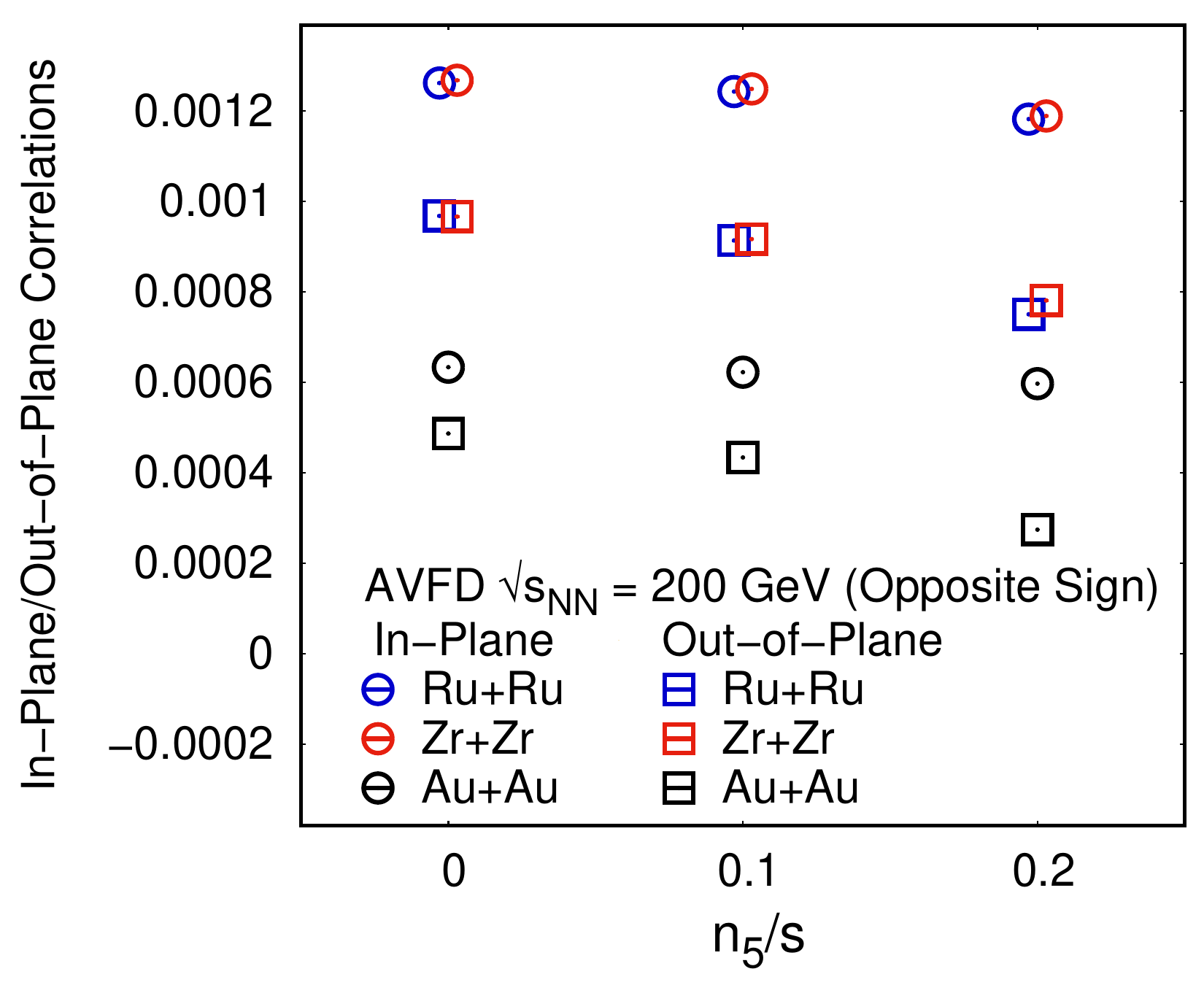}
  \includegraphics[width=.44\textwidth]{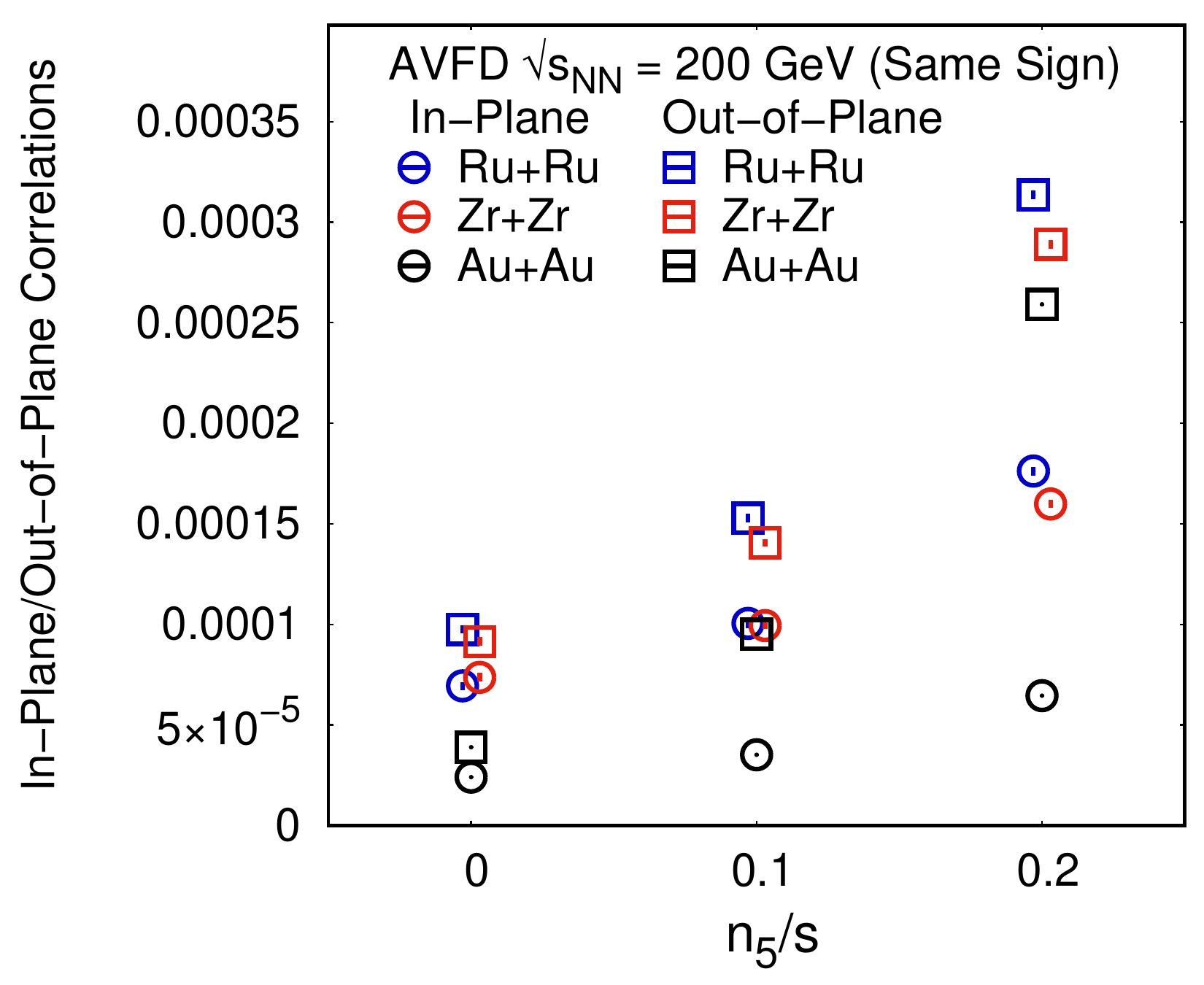}
  \caption{(Color Online) In-plane and out-of-plane correlations for opposite sign (left) and same sign (right) charge pairs versus $n_5/s$ for the 30-40$\%$ collision centrality. Markers are slightly shifted along the x-axis for clarity. Statistical uncertainties are small and are within the marker size.}
  \label{fig:Inplane}
\end{figure*}
\begin{figure*}[htbp]
  \centering
  \includegraphics[width=.90\textwidth]{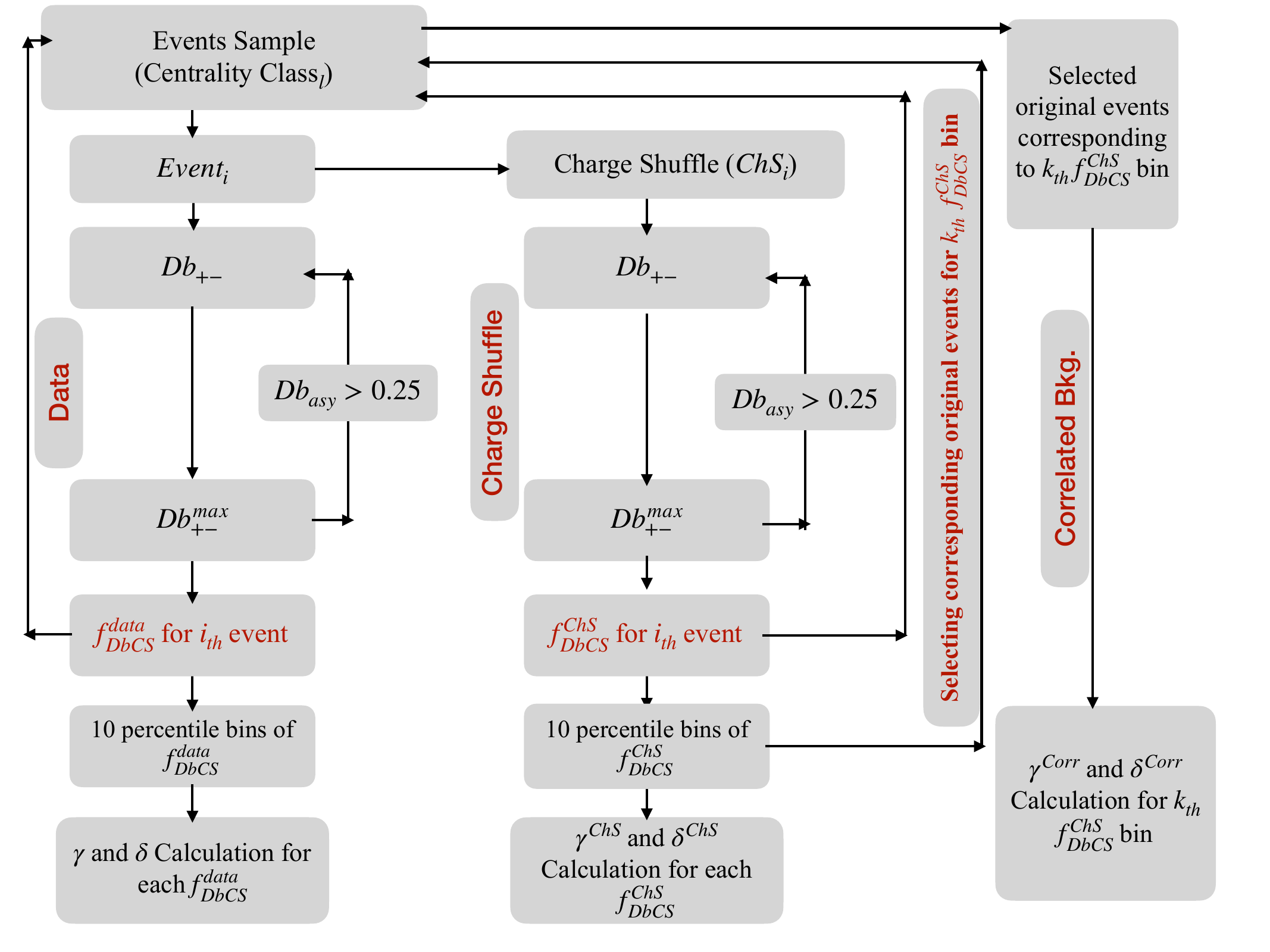}
  \caption{(Color Online) Flow chart displaying various steps involved in computing $\gamma$ and $\delta$ correlators employing the SDM.}
  \label{fig:flowchart}
\end{figure*}

\section{Results and discussion}
\label{sec:Results}
Figure~\ref{fig:gamma} (top left) shows the $\gamma$-correlators for opposite-sign (OS) and same-sign (SS) charge pairs in $Au+Au$ and isobar ($Ru+Ru$ and $Zr+Zr$) collisions at $\sqrt{s_{\mathrm NN}}$ = 200 GeV versus axial charge per entropy density. The results indicate that $\gamma$ is negative for SS charge pairs and positive for OS charge pairs. The magnitude of $\gamma$ increases as $n_5/s$ increases from $0.0$ to $0.2$. Additionally, the $\gamma$ values for the two isobar collisions ($Ru+Ru$ and $Zr+Zr$) are similar within errors for both SS and OS charge pairs.
\begin{figure*}
  \centering
  \includegraphics[width=.475\textwidth]{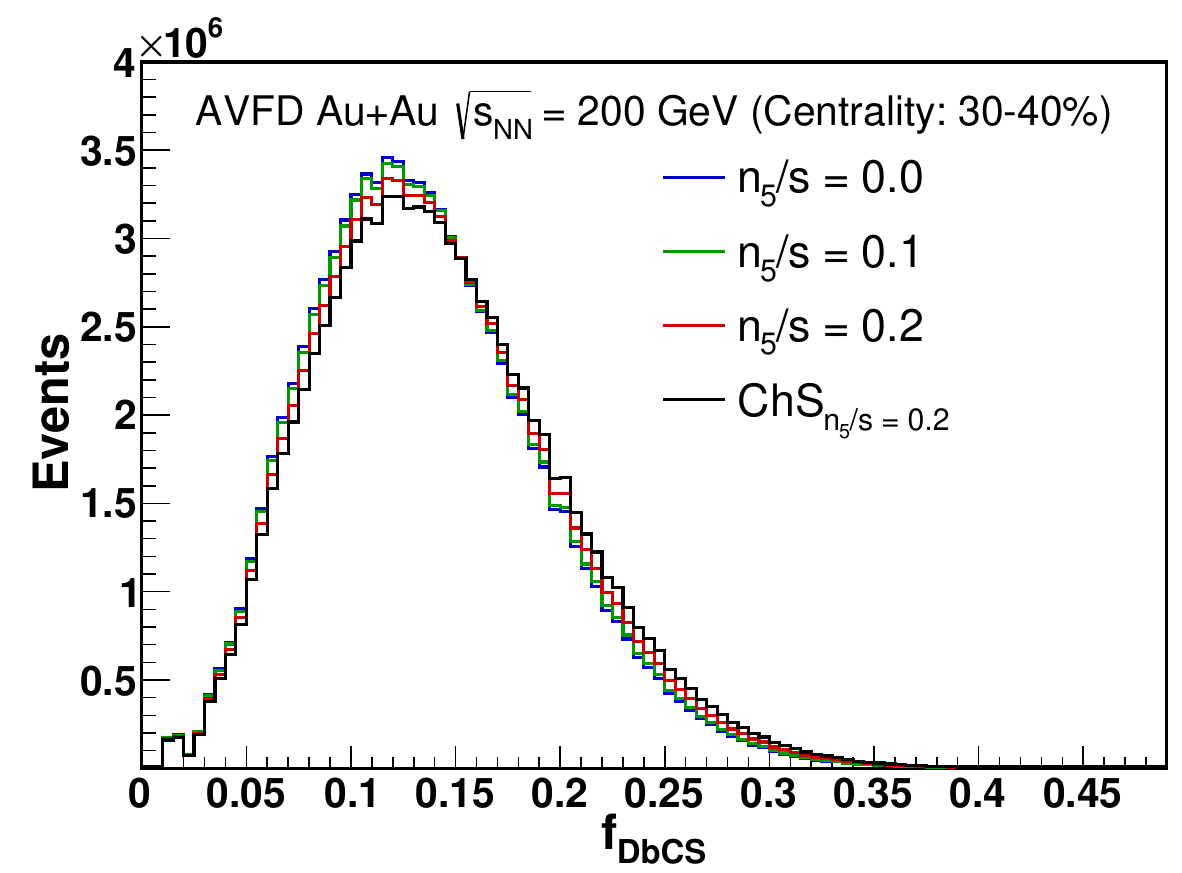}
  \includegraphics[width=.475\textwidth]{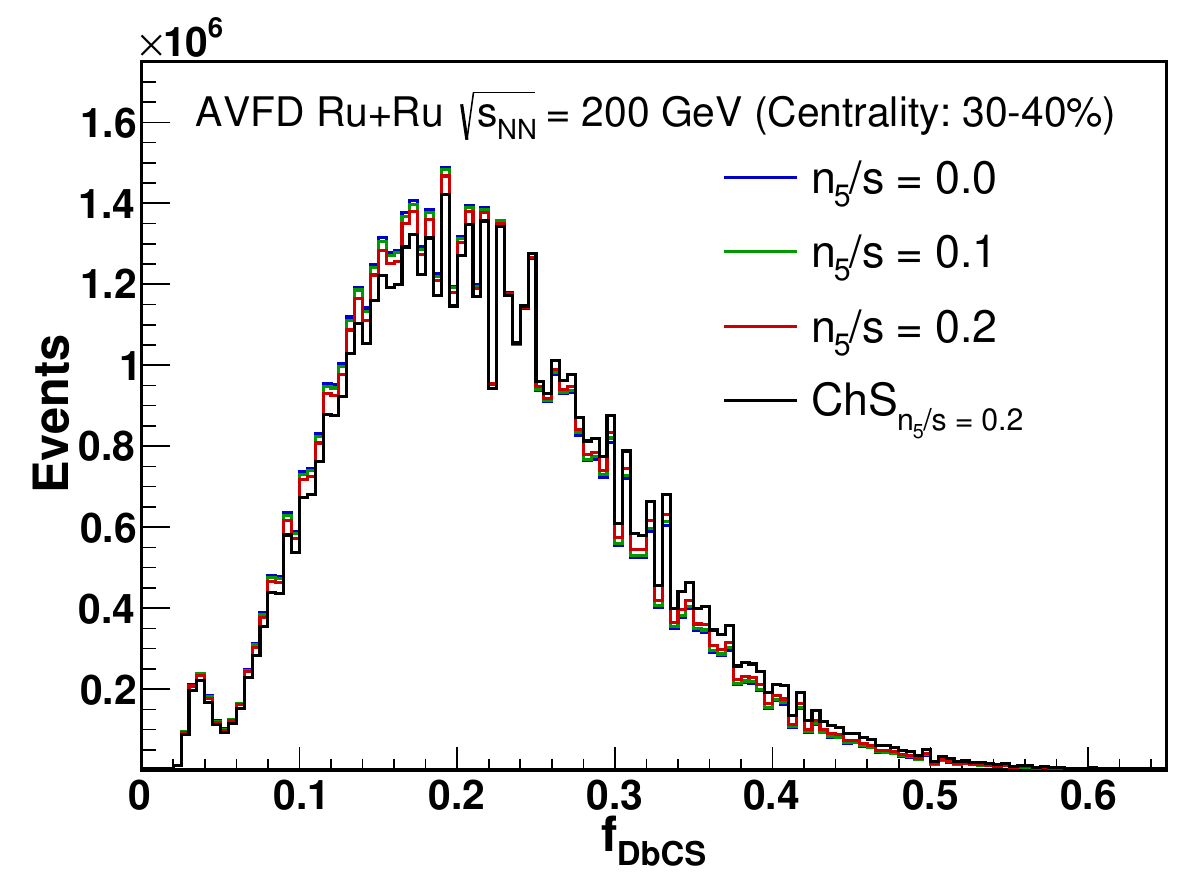}
  \caption{$f_{DbCS}$ distributions for AVFD generated $Au+Au$ (left) and $Ru+Ru$ (right) collisions at $\sqrt{s_{\mathrm NN}}$ = 200 GeV. The rightmost side of the distribution represents the highest charge separation (0-10$\%$ $f_{DbCS}$) and the leftmost side of the distribution represents the lowest charge separation (90-100$\%$ $f_{DbCS}$).}
  \label{fig:fDBCS_dist}
\end{figure*}
Notably, $\gamma$ is larger for OS pairs in isobar collisions compared to $Au+Au$ collisions, which is due to the increased background associated with the lower multiplicities in isobar collisions. The reaction plane independent $\delta$-correlators for isobar and $Au+Au$ collisions for different $n_5/s$ values, are displayed in the Fig.~\ref{fig:gamma} (top right).
\begin{figure*}[htbp]
  \centering
  \includegraphics[width=.46\textwidth]{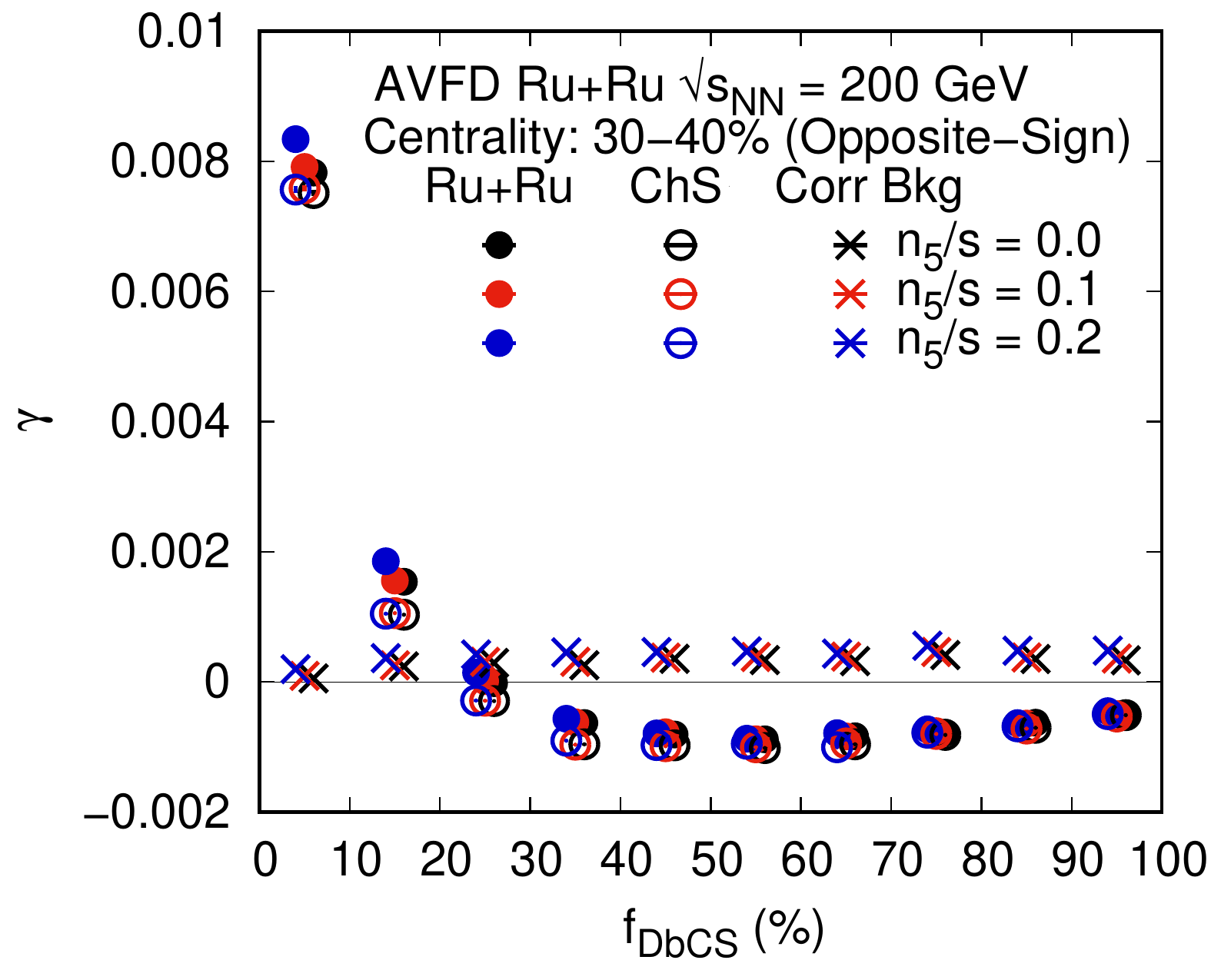}
  \includegraphics[width=.46\textwidth]{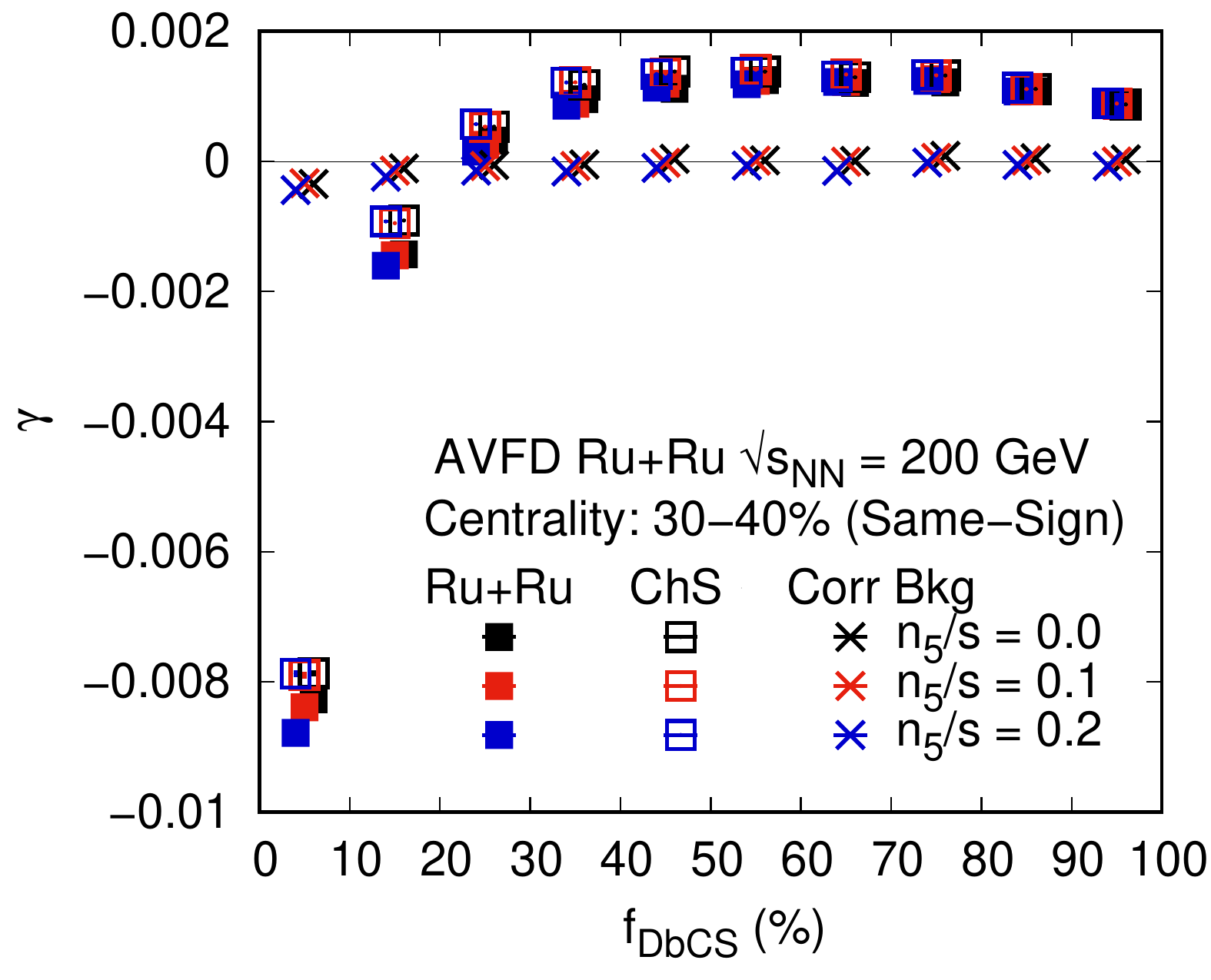}
  
  \includegraphics[width=.46\textwidth]{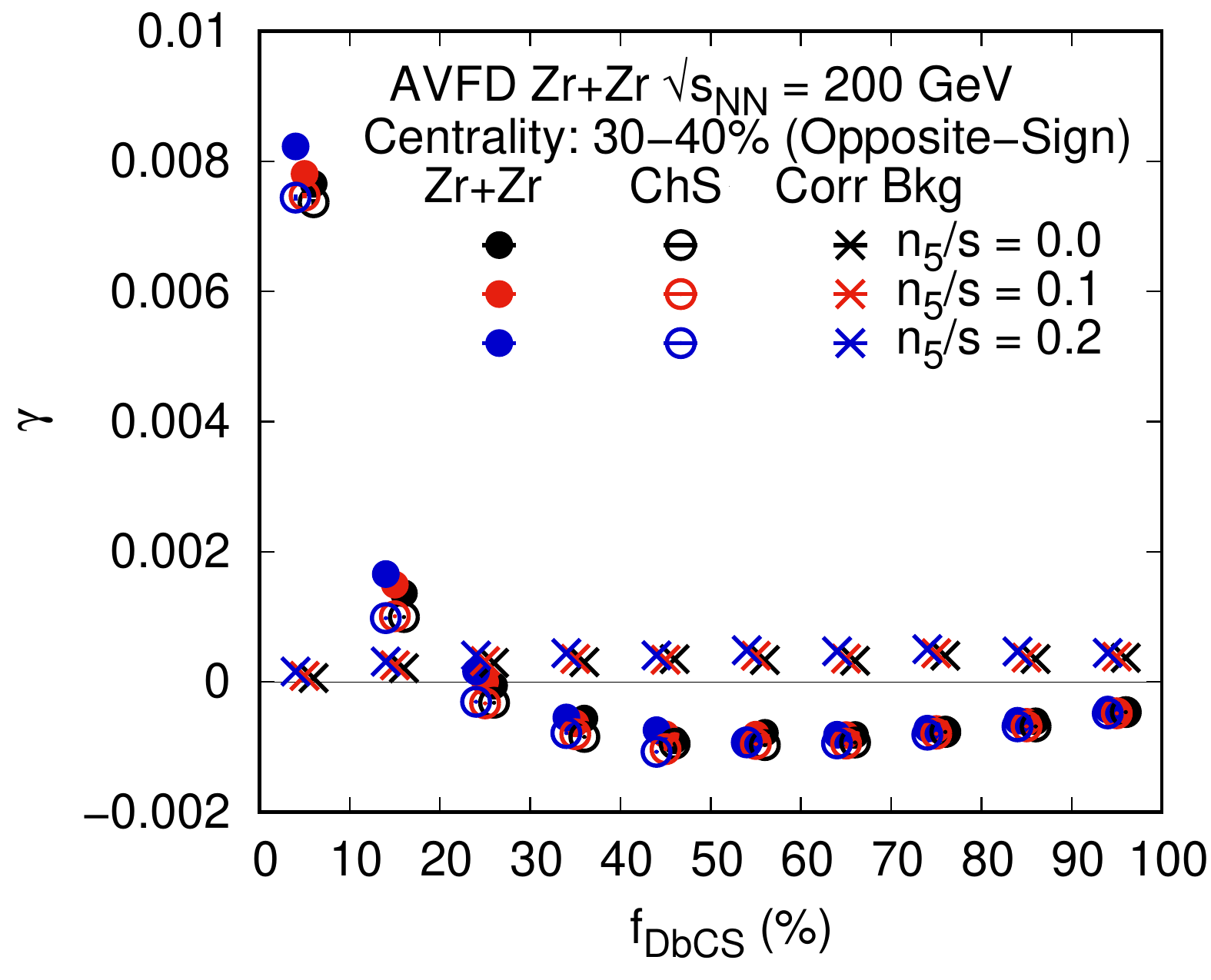}
  \includegraphics[width=.46\textwidth]{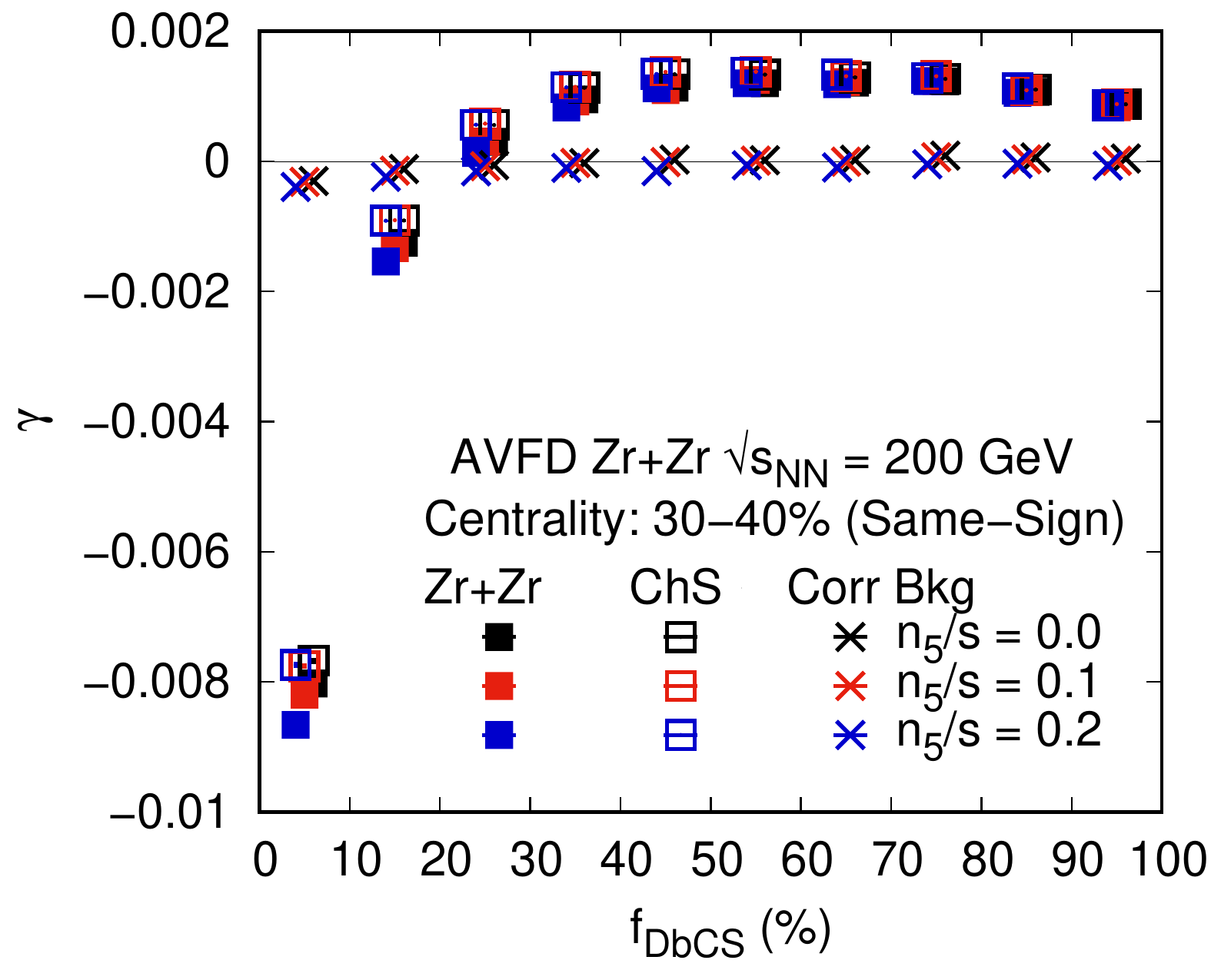}
  
  \includegraphics[width=.46\textwidth]{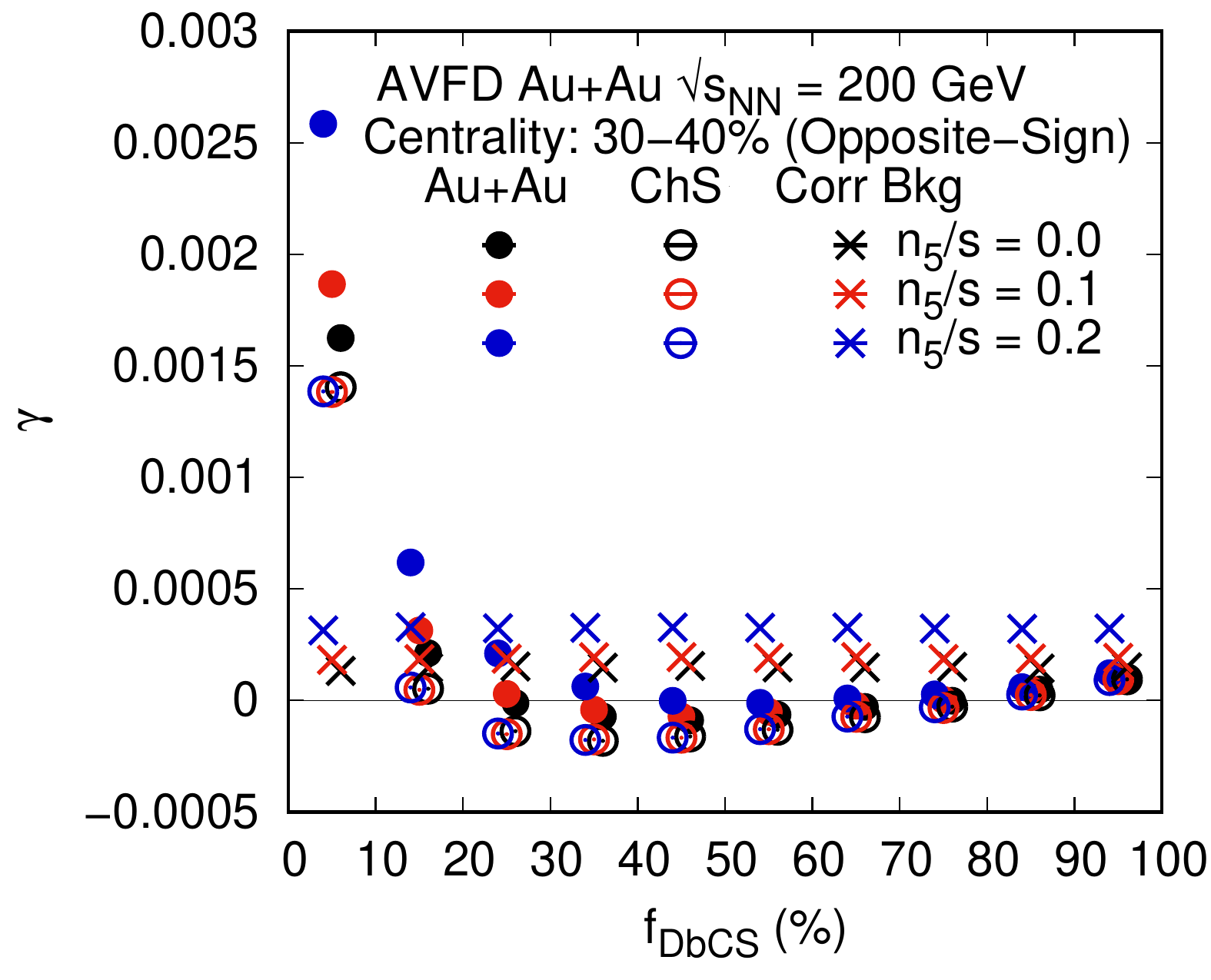}
  \includegraphics[width=.46\textwidth]{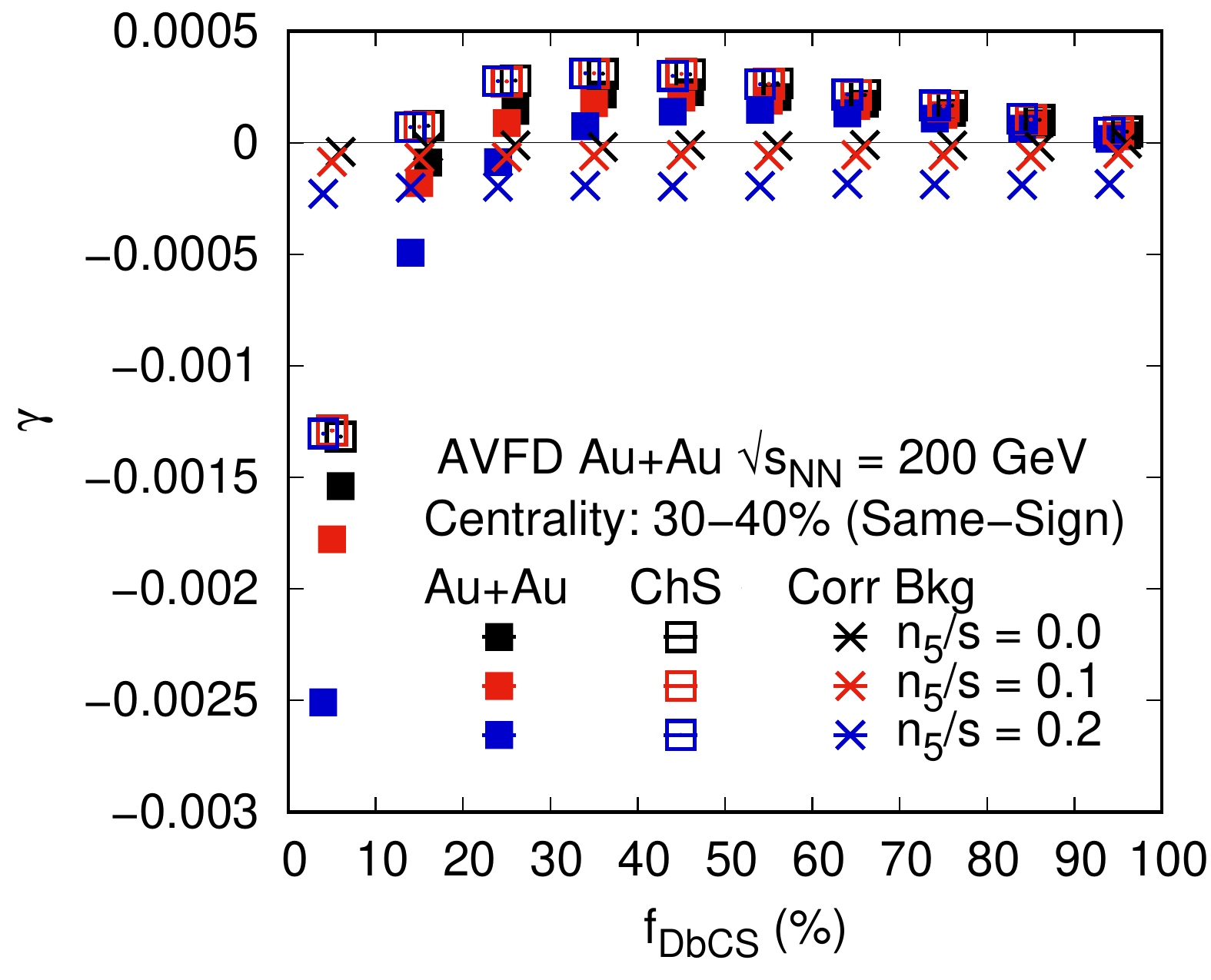}
  \caption{(Color online) $\gamma$-correlators as a function of $f_{DbCS}$ for $Ru+Ru$ (top), $Zr+Zr$ (middle) and $Au+Au$ (bottom) collisions at $\sqrt{s_{\mathrm NN}}$ = 200 GeV for opposite-sign (Left) and  same-sign (Right), for three CME samples. $\gamma$-correlators for charge-shuffle ($\gamma_{ChS}$) and correlated ($\gamma_{Corr}$) backgrounds  are also shown. The markers are slightly shifted along the x-axis for clarity. Statistical uncertainties are small and are within the marker size.}
  \label{fig:gamma_fdbcs}
\end{figure*}
Both OS and SS charge pairs have positive $\delta$ values, but the OS pairs exhibit larger values. It is noteworthy that the values of $\gamma$ and $\delta$ obtained in the present analysis are consistent with those reported in Ref.~\cite{avfd_star}. Again, it is observed that $\delta$ values are higher in isobar collisions compared to those in $Au+Au$ collisions, attributed to the increased background in the former. Figure~\ref{fig:gamma} (Bottom) shows the dependence of the CME-sensitive $\Delta\gamma$ on $n_5/s$ for isobar and $Au+Au$ collisions for 30-40$\%$ collision centrality. The data points at $n_5/s=0.0$ show significant $\Delta\gamma$ values, despite the expectation of near-zero values in the absence of a CME signal. This anomaly is due to 33$\%$ LCC, which mimics a CME signal. However, in AMPT generated Au+Au collisions at $\sqrt{s_{\mathrm NN}}$ = 200 GeV without 33$\%$ LCC, $\Delta\gamma$ was found to be approximately zero~\cite{Aggarwal_SDM_2022}. The $\Delta\gamma$ at $n_5/s=0$ is almost twice in isobar collisions compared to $Au+Au$ collisions which scales inversely with multiplicities~\cite{Lacey:2022plw}. The multiplicities in isobar collisions are approximately half compared to those in $Au+Au$ collisions. The relative increase in $\Delta\gamma$ from $n_5/s=0$ to $n_5/s=0.1$ and 0.2 is more pronounced in $Au+Au$ collisions than in isobar collisions.\par
In-plane and out-of-plane correlations for the opposite- and same-sign charged pairs are displayed, respectively, in Fig.~\ref{fig:Inplane} (left) and Fig.~\ref{fig:Inplane} (right). Both in-plane and out-of-plane correlations are found to be positive for both OS and SS charge pairs. The OS charge pairs show stronger in-plane correlations whereas SS charge pairs exhibit stronger out-of-plane correlations.

\subsection{Analyzing data using Sliding Dumbbell Method}
\label{sec:usingSDM}
\begin{figure*}[htbp]
  \centering
  \includegraphics[width=.45\textwidth]{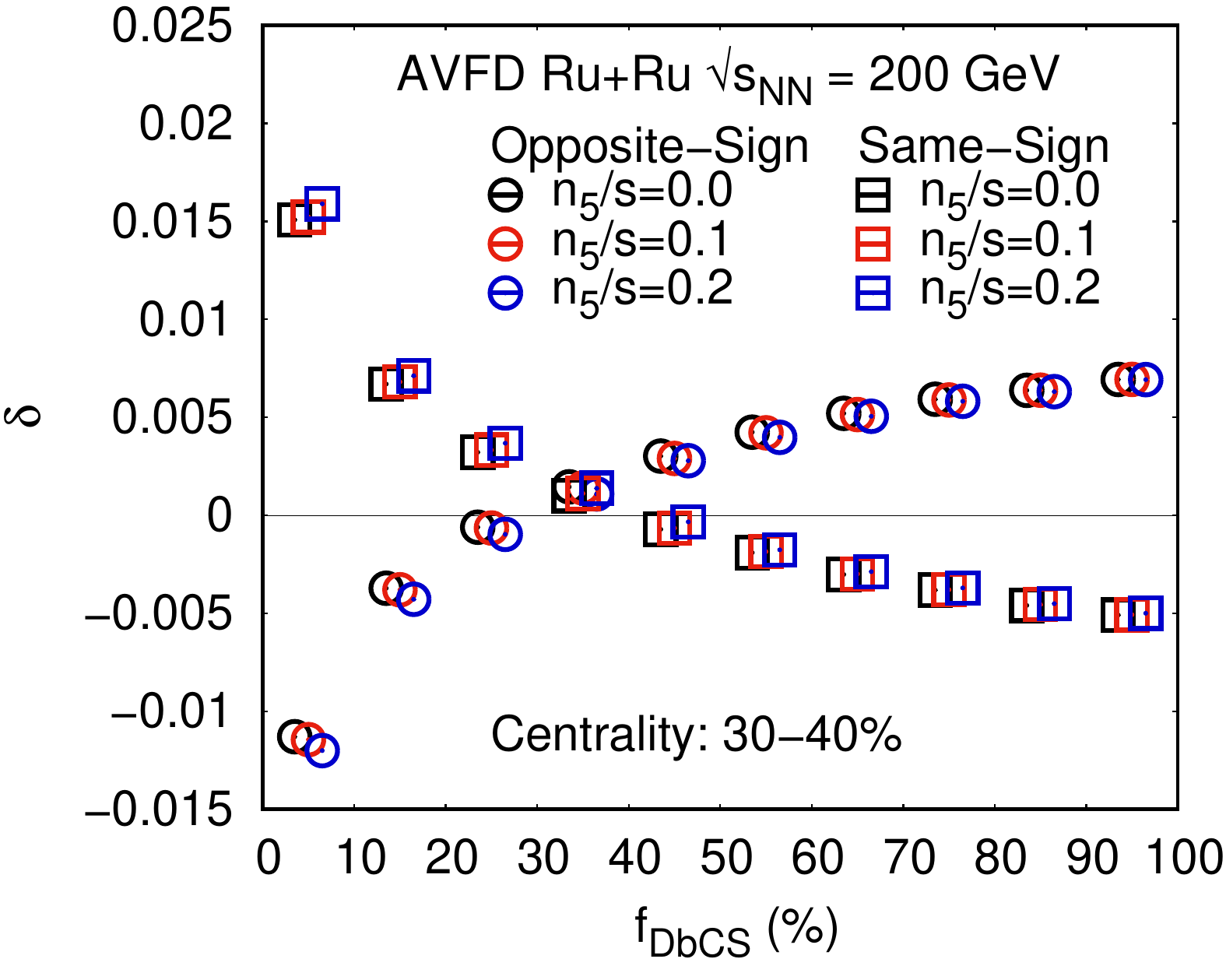} 
  \includegraphics[width=.45\textwidth]{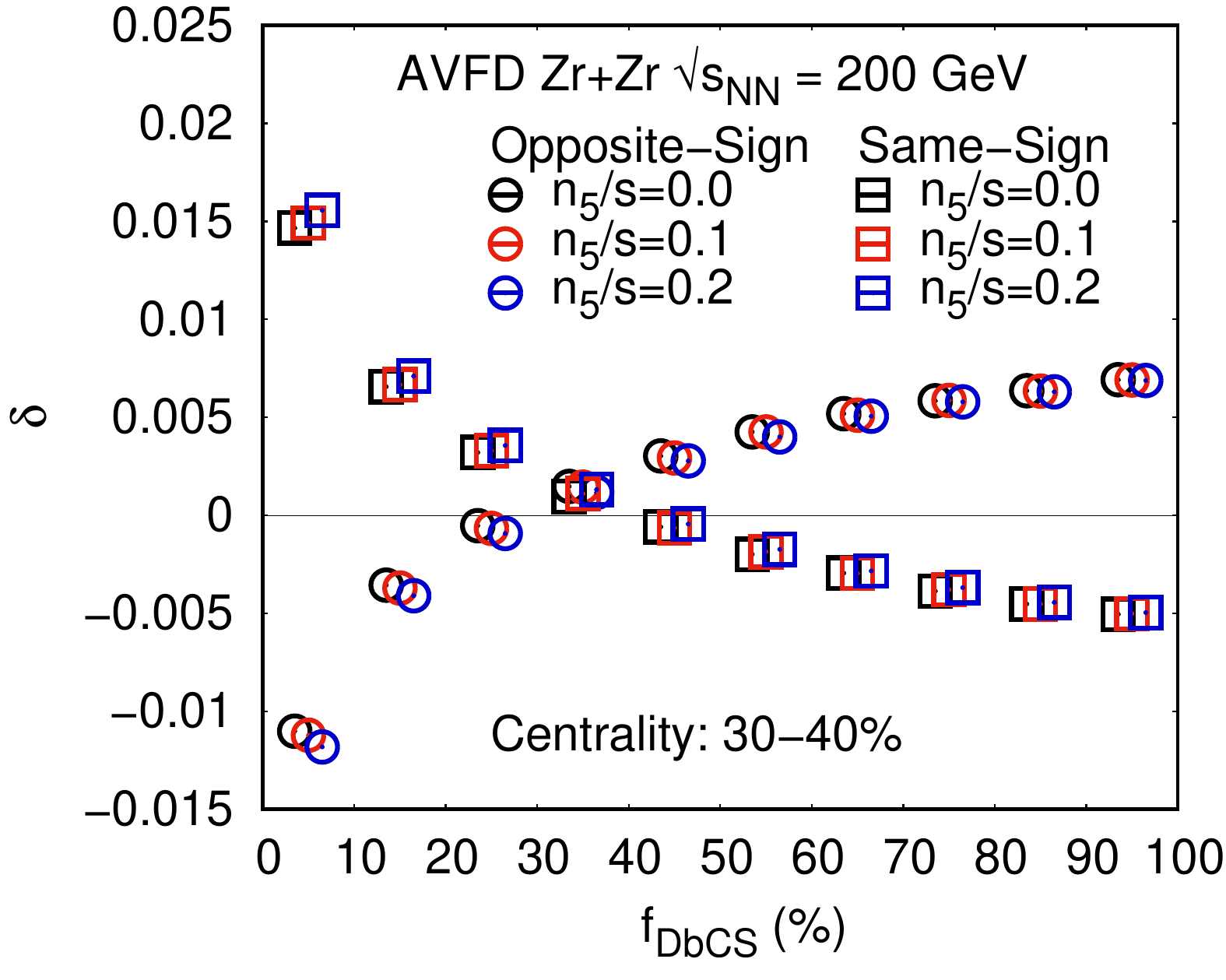}
  \includegraphics[width=.45\textwidth]{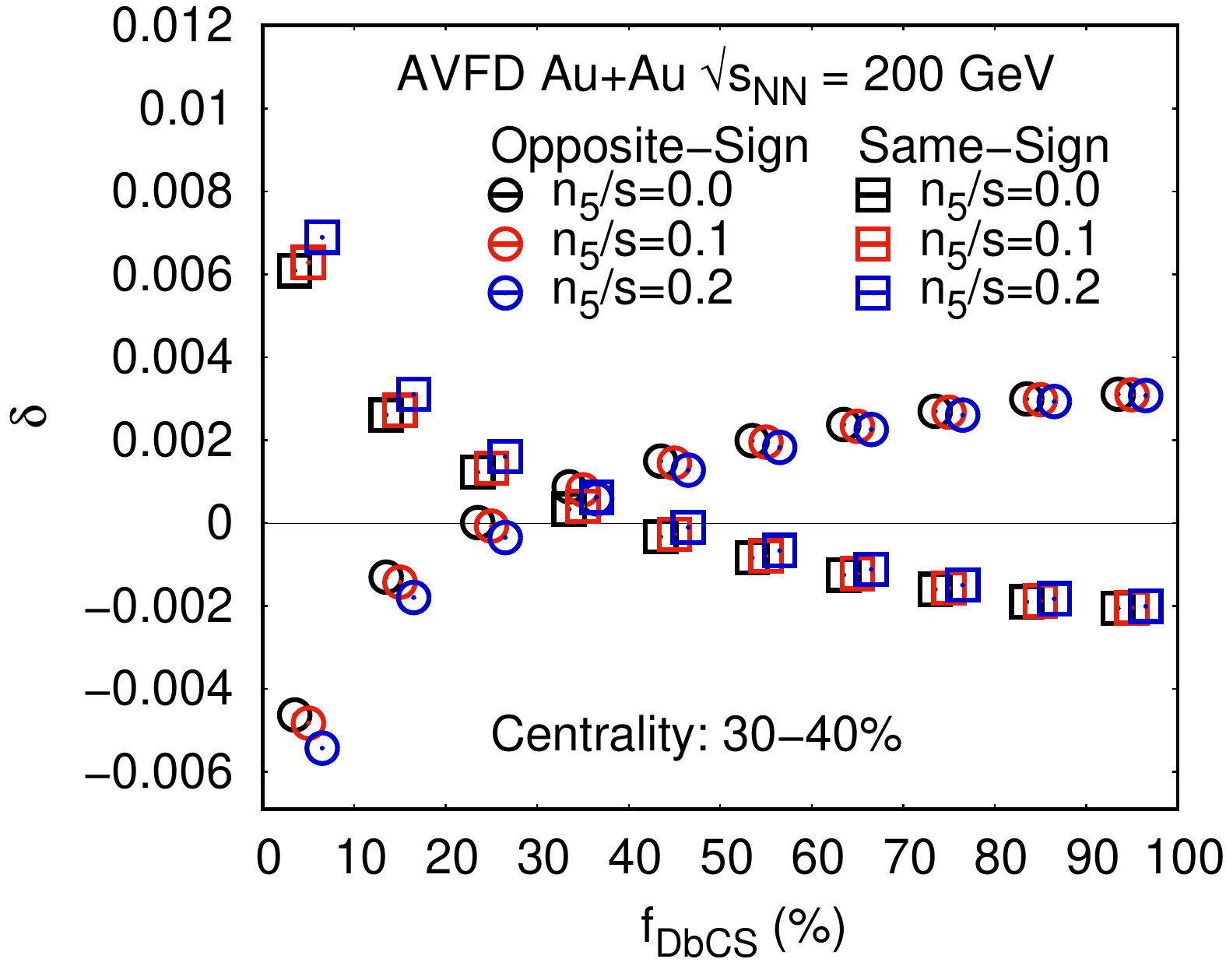} 
  \caption{(Color online) $\delta$-correlators as a function of $f_{DbCS}$ for $Ru+Ru$ (top left), $Zr+Zr$ (top right) and $Au+Au$ (bottom) collisions at $\sqrt{s_{\mathrm NN}}$ = 200 GeV for opposite-sign and same-sign charge pairs for different CME samples. Markers are slightly shifted along the x-axis for clarity. Statistical uncertainties are small and are within the marker size.}
  \label{fig:delta11}
\end{figure*}
Figure~\ref{fig:flowchart} displays the flow chart describing the various steps involved in the analysis as discussed in sections~\ref{sec:SDM} and~\ref{sec:DataAna}. In the initial phase of the analysis, the entire azimuthal plane of each event is scanned, and the $f_{DbCS}$ distributions are computed for the 30-40$\%$ collision centrality and for different data sets as listed in Table~\ref{tab:avfd} and their corresponding charge shuffle samples. These distributions are subsequently categorized into ten percentile bins, ranging from 0–10$\%$ (representing the highest charge separation) to 90–100$\%$ (representing the lowest charge separation). Following this, multi-particle correlators (2-, 3-, and 4-particle) are calculated for each $f_{DbCS}$ bin, utilizing samples from AVFD, charge shuffle, and correlated backgrounds.\par
\begin{figure*}[htbp]
  \centering
  \includegraphics[width=.42\textwidth]{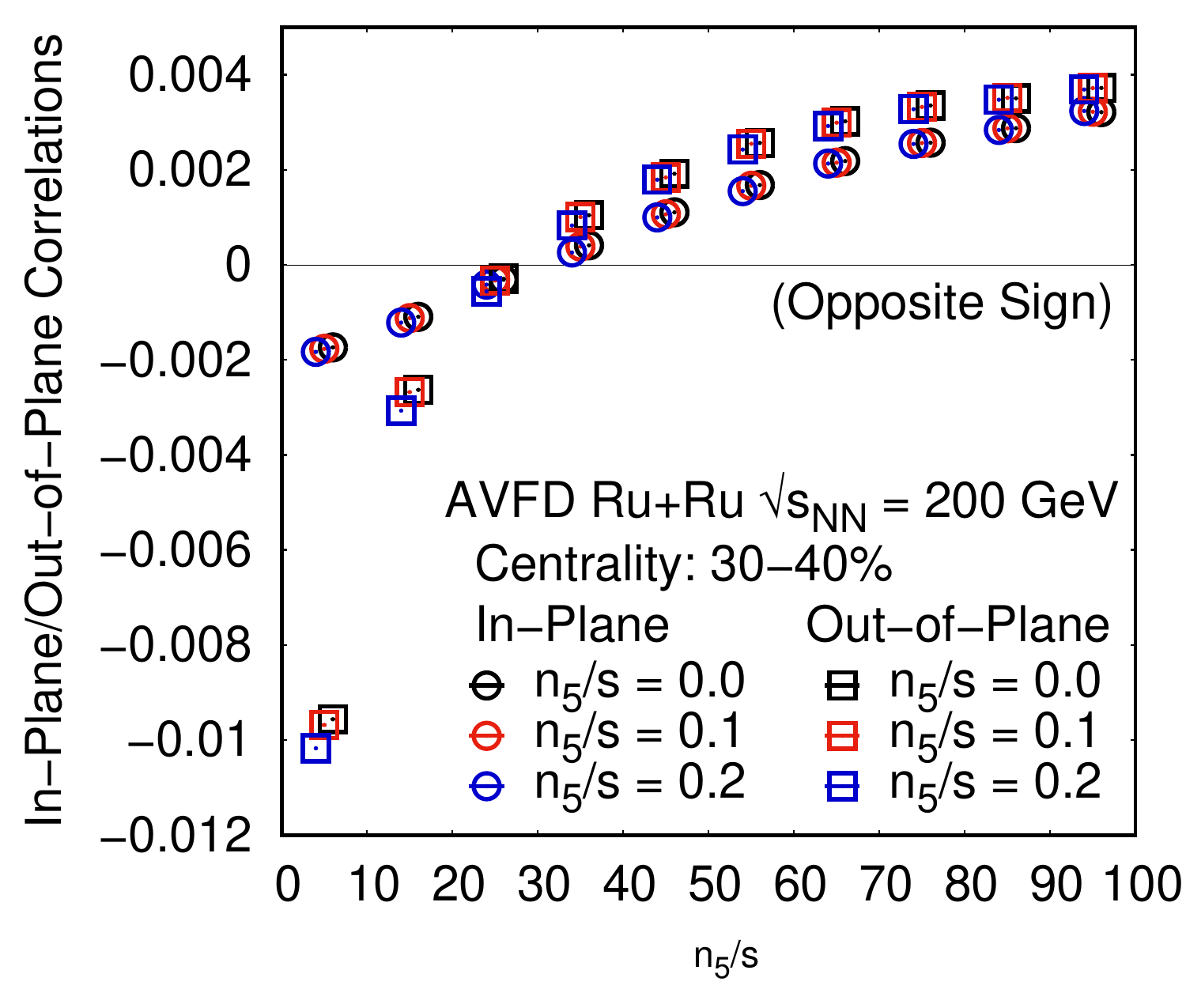}
  \includegraphics[width=.42\textwidth]{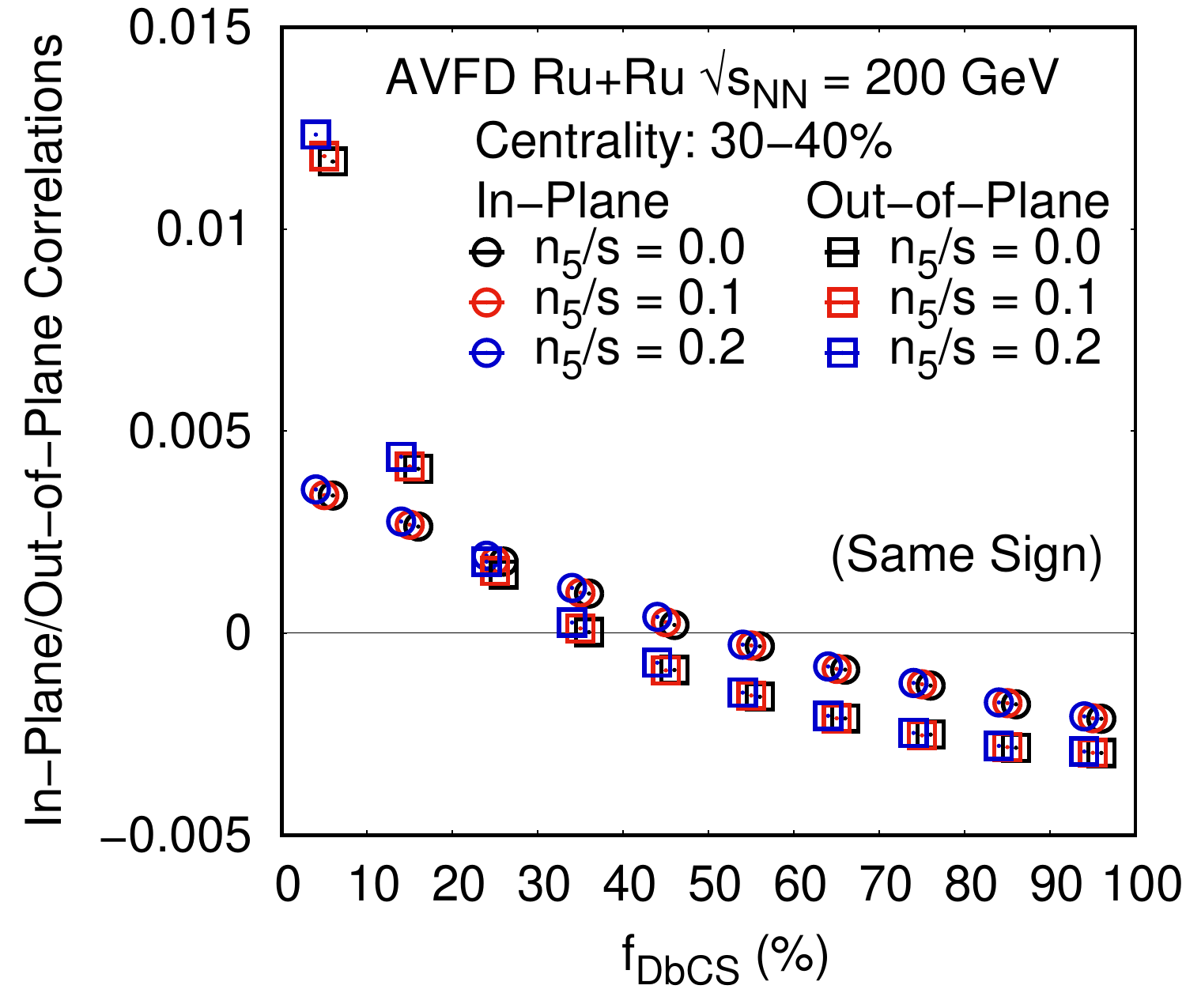}
  \includegraphics[width=.42\textwidth]{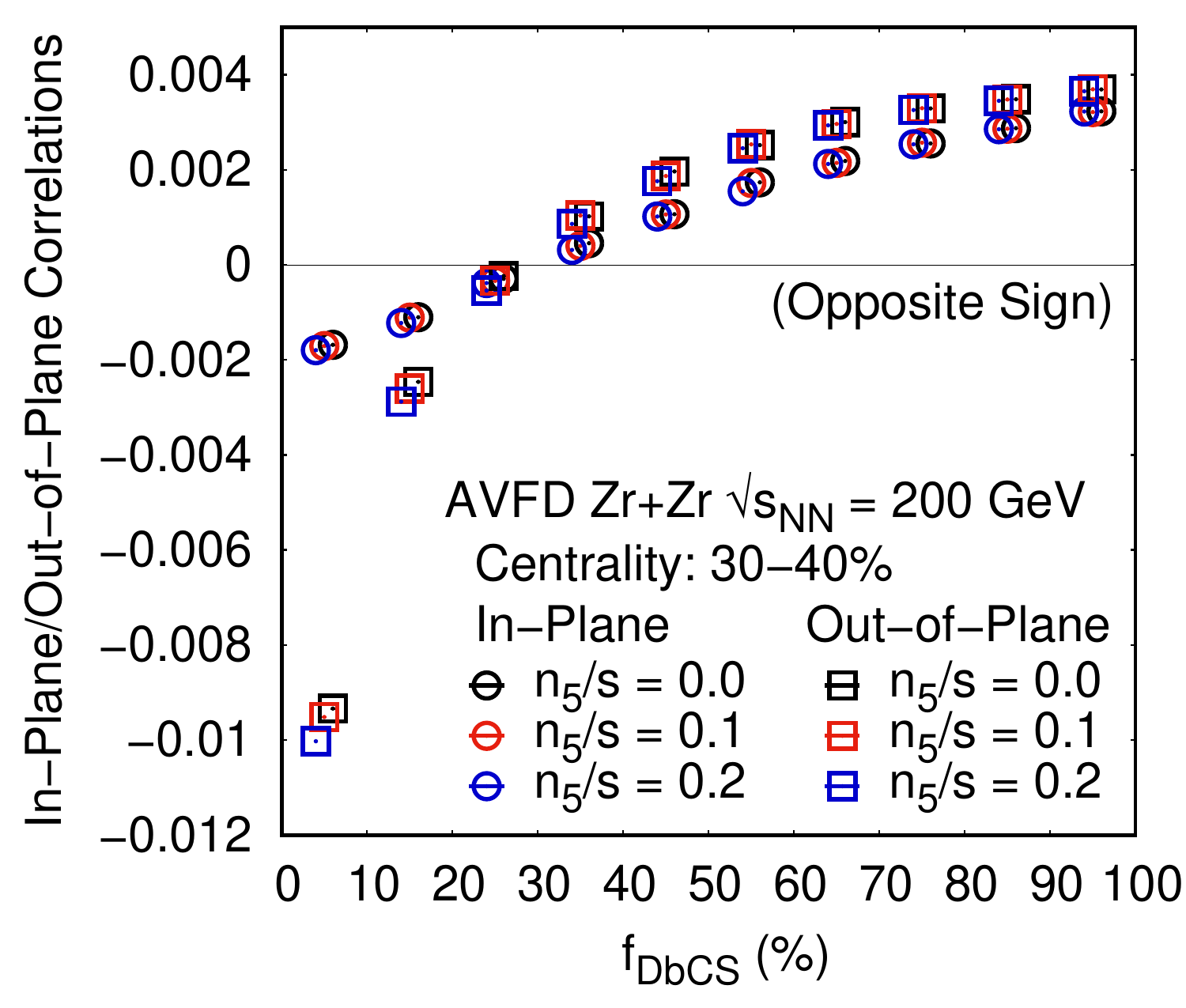}
  \includegraphics[width=.42\textwidth]{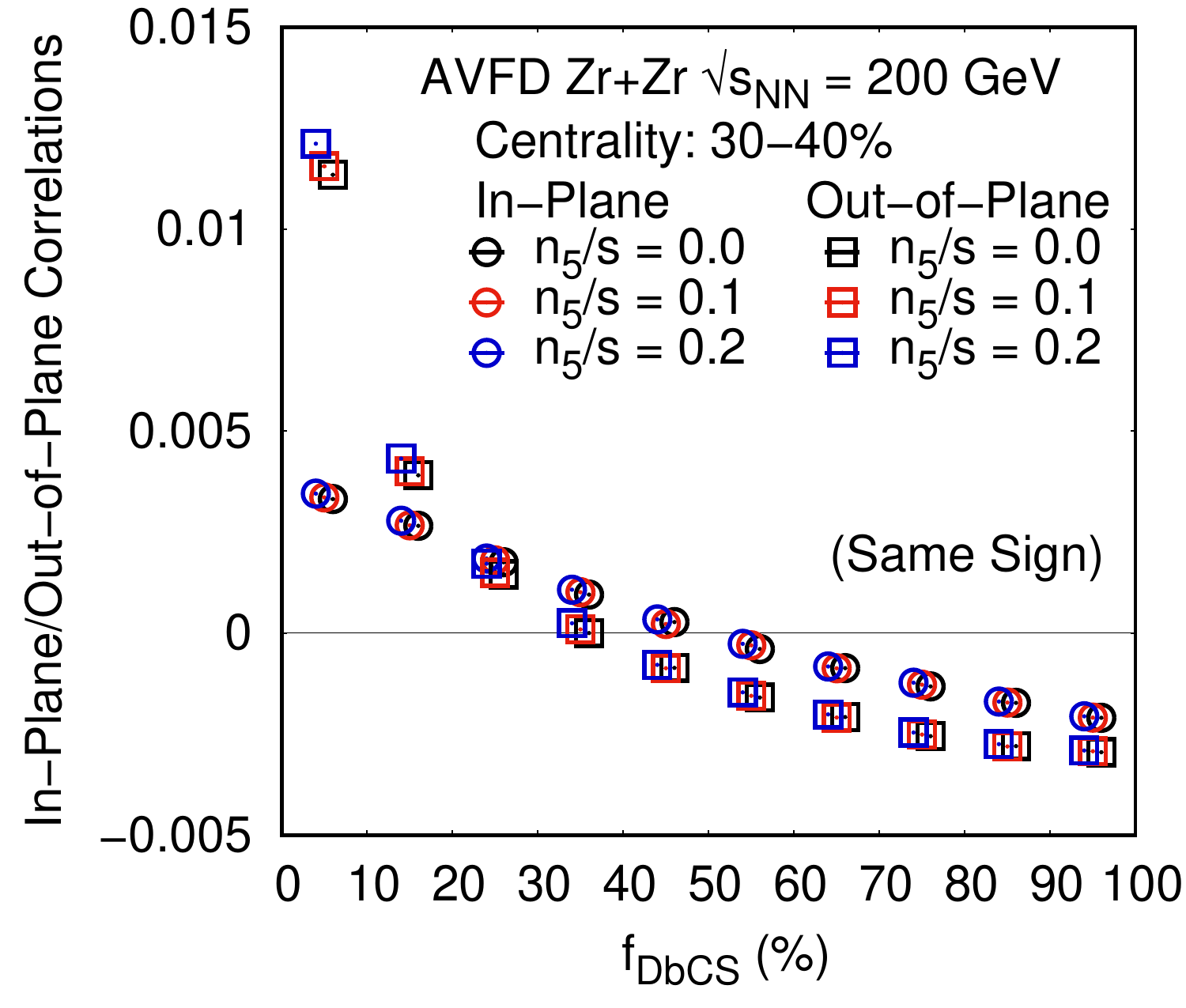}
  \includegraphics[width=.42\textwidth]{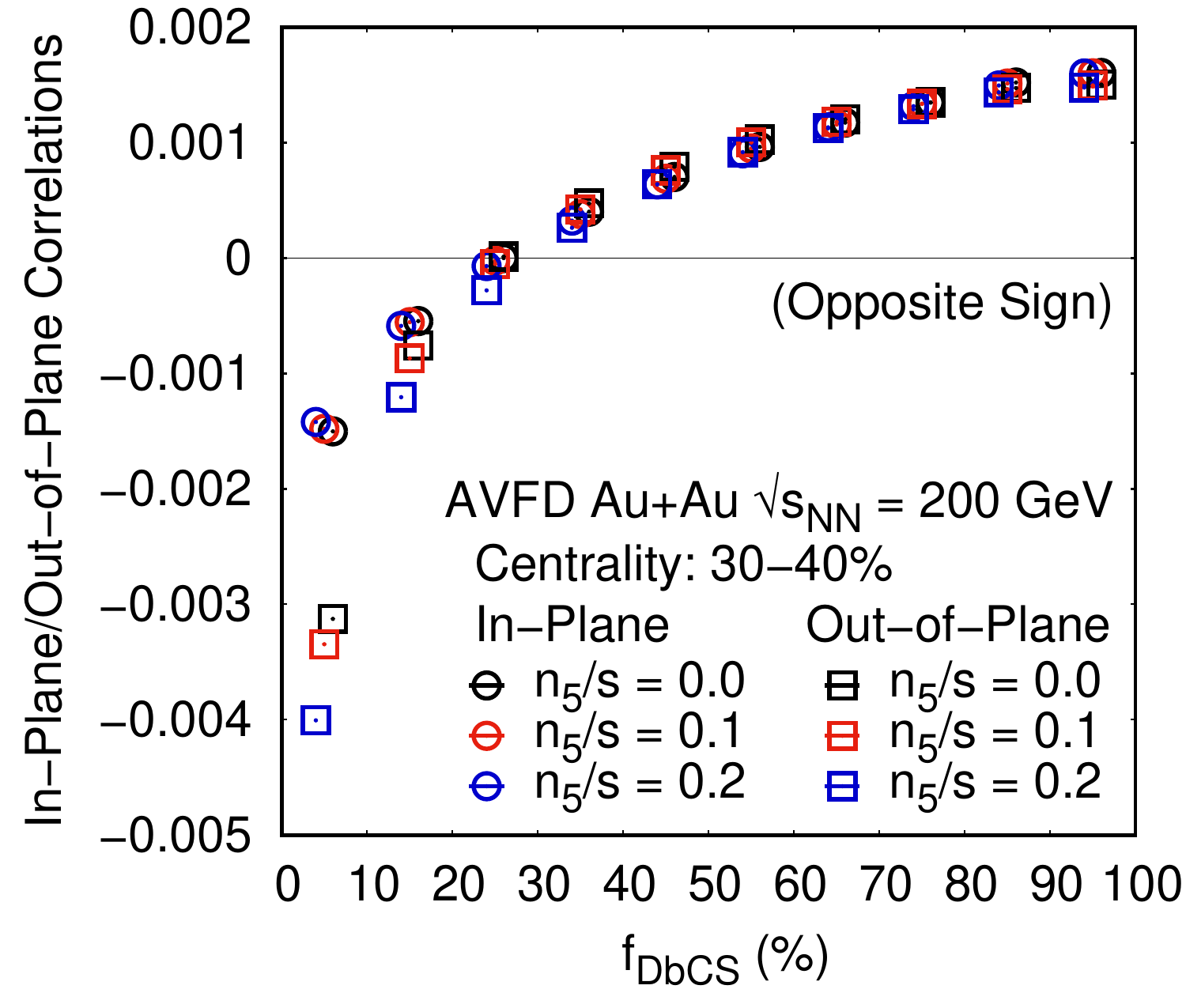}
  \includegraphics[width=.42\textwidth]{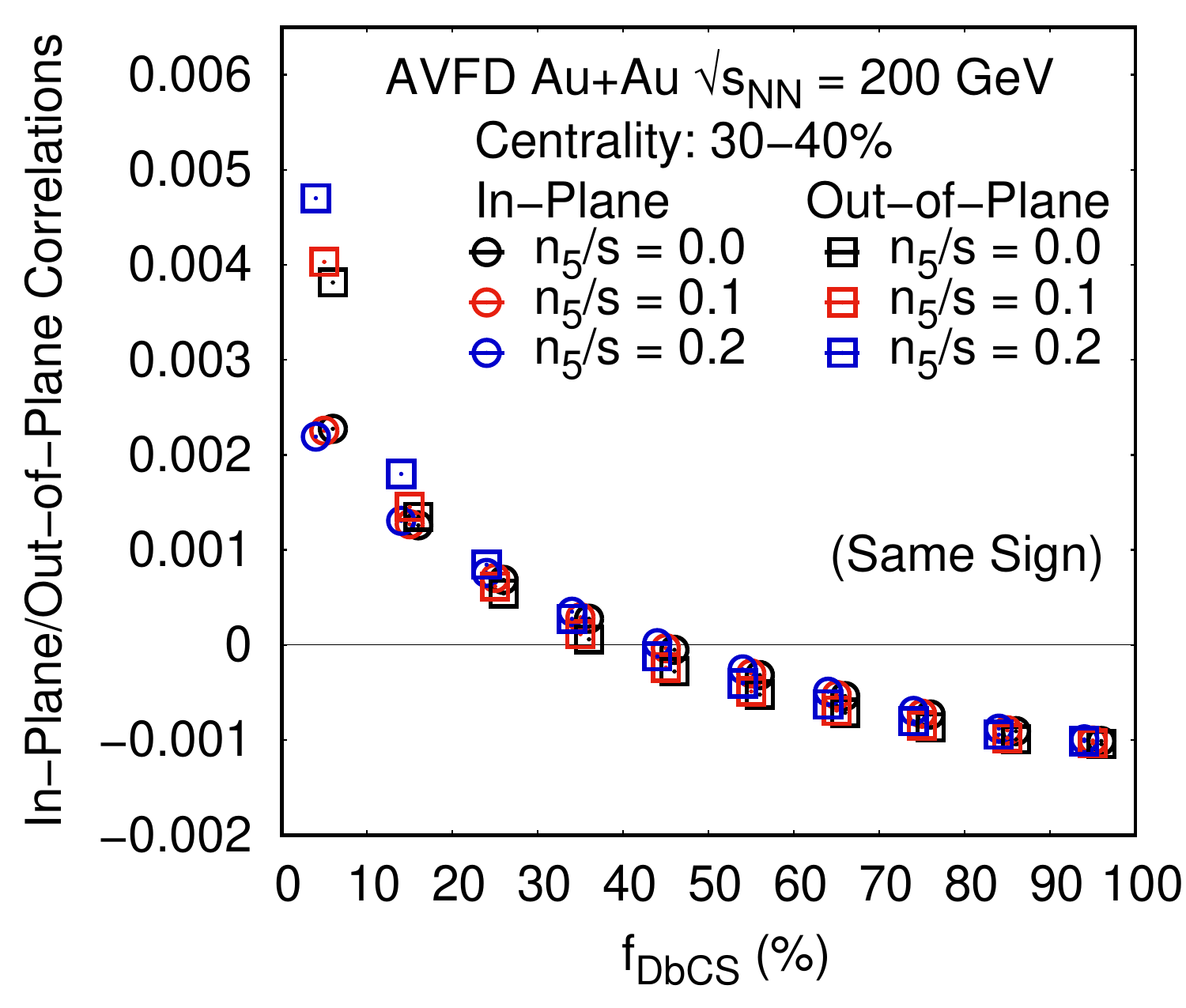}
  \caption{(Color online) In-pane and out-of-plane correlations as a function of $f_{DbCS}$ for $Ru+Ru$ (top), $Zr+Zr$ (middle) and $Au+Au$ (bottom) collisions at $\sqrt{s_{\mathrm NN}}$ = 200 GeV for opposite-sign (Left) and  same-sign (Right), for different CME samples. Markers are slightly shifted along the x-axis for clarity. Statistical uncertainties are small and are within the marker size.}
  \label{fig:InplanefDbCS}
\end{figure*}
The $f_{DbCS}$ distributions for $Au+Au$ and $Ru+Ru$ collisions at $\sqrt{s_{\mathrm NN}}$ = 200 GeV, corresponding to different axial charge per entropy density ($n_5/s$), are shown in Fig.~\ref{fig:fDBCS_dist} (left) and Fig.~\ref{fig:fDBCS_dist} (right), respectively. The $f_{DbCS}$ distributions for the charge shuffle (ChS) samples are also displayed in figures for comparison. For $Zr+Zr$ collisions, $f_{DbCS}$ distributions similar to $Ru+Ru$ collisions have been observed. These distributions show slight forward shift along with decreasing peak with increasing $n_5/s$ values. Furthermore, the $f_{DbCS}$ distributions for the charge shuffle  across the various $n_5/s$ values are nearly indistinguishable, so only distributions for $n_5/s$ = 0.2 are shown. These $f_{DbCS}$ distributions are divided into 10 percentile bins as discussed above. This method of partitioning events based on $f_{DbCS}$ helps to identify potential CME-like events characterized by the highest back-to-back charge separation across the dumbbell.\par

\begin{figure*}
  \centering %
  \includegraphics[width=.42\textwidth]{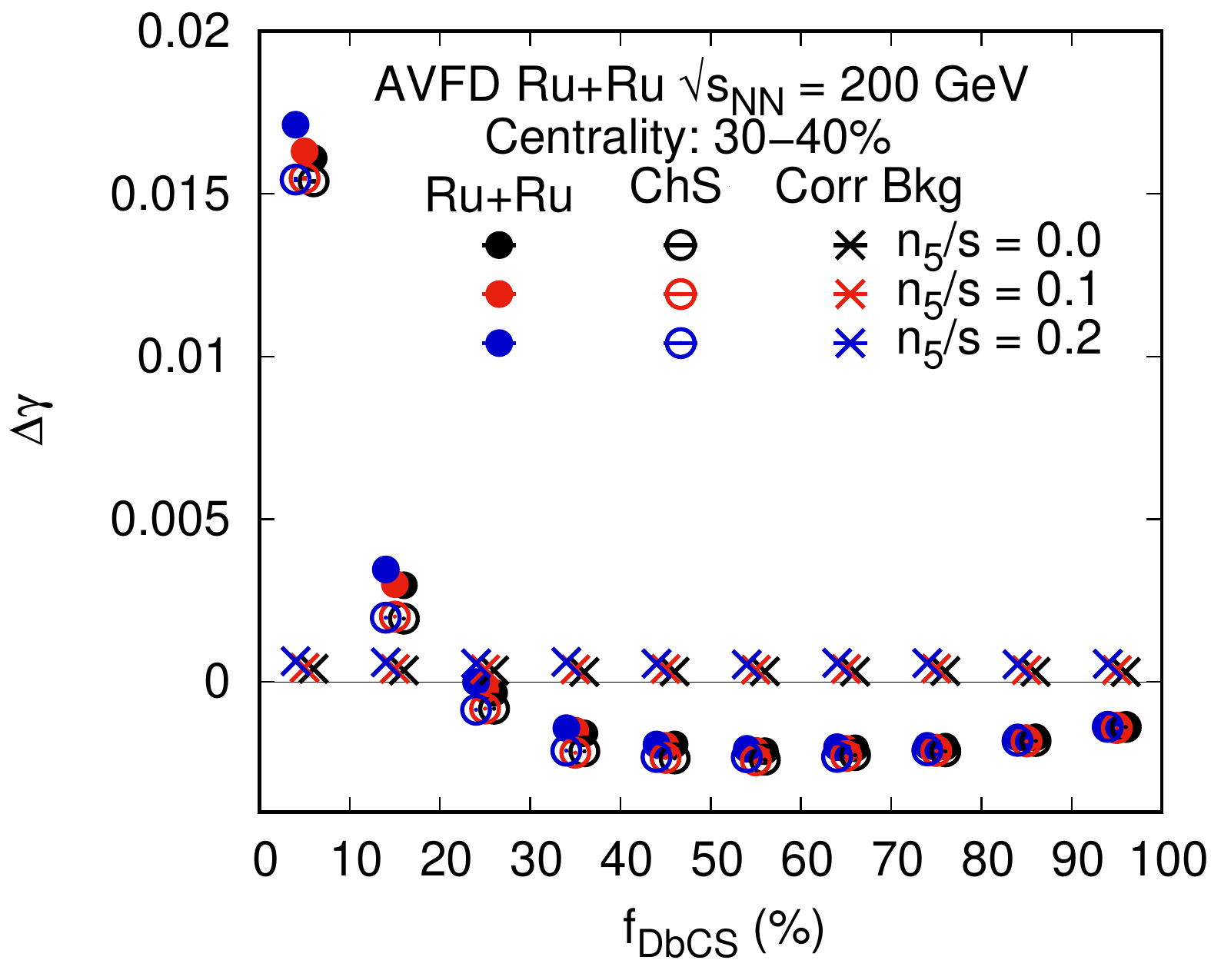}
  \includegraphics[width=.42\textwidth]{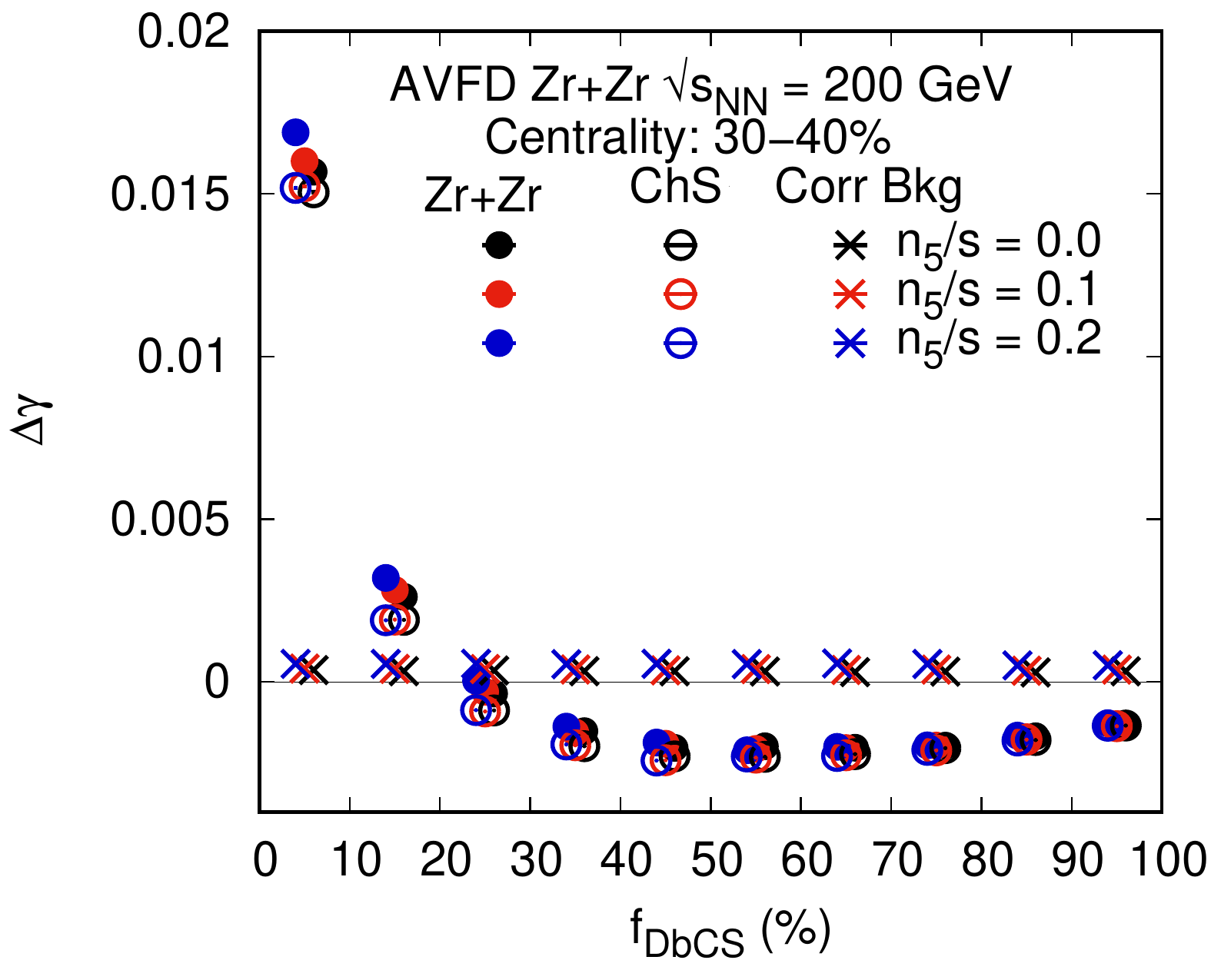}
  \includegraphics[width=.42\textwidth]{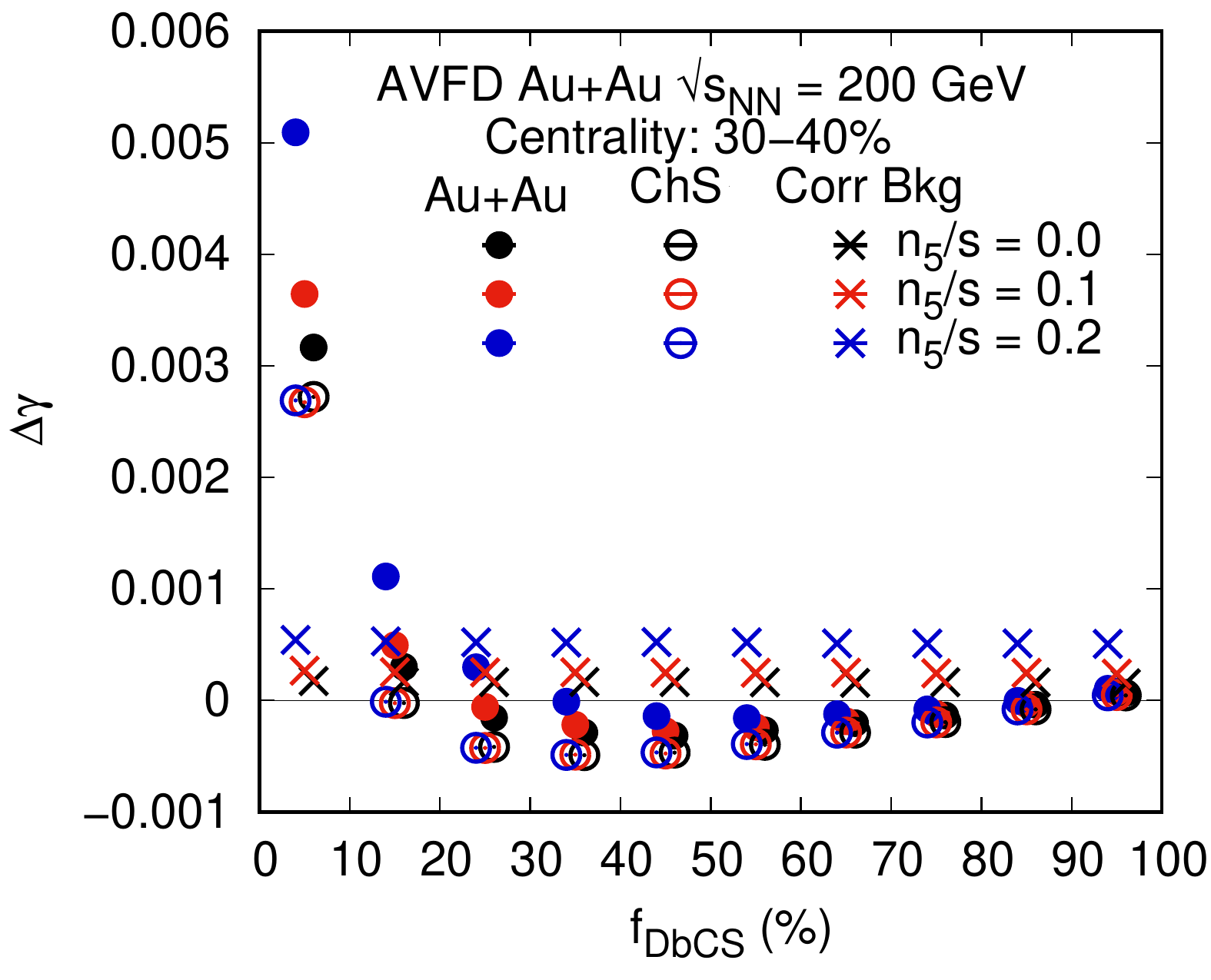}
  \caption{(Color online) $\Delta\gamma$ (=$\gamma_{OS}$-$\gamma_{SS}$) as a function of $f_{DbCS}$ for AVFD generated $Ru+Ru$ (top left), $Zr+Zr$ (top right) and $Au+Au$ (bottom) collisions at $\sqrt{s_{\mathrm NN}}$ = 200 GeV, for different CME samples. $\Delta\gamma$ for charge shuffle background ($\gamma_{ChS}$) and correlated background ($\gamma_{Corr}$) are also displayed as open circles and cross markers, respectively. Markers are slightly shifted along the x-axis for clarity. Statistical uncertainties are small and are within the marker size.}
  \label{fig:Deltagamma_fdbcs}
\end{figure*}
Figure~\ref{fig:gamma_fdbcs} displays the $\gamma$-correlators for opposite-sign (left) and same-sign (right) charge pairs as a function of $f_{DbCS}$ percentile bins for $Ru+Ru$ (top), $Zr+Zr$ (middle), and $Au+Au$ (bottom) collisions. The comparisons to charge shuffle ($\gamma_{ChS}$) and correlated ($\gamma_{Corr}$) backgrounds  are also shown. The magnitude of the $\gamma$-correlators increases for both same-sign (SS) and opposite-sign (OS) charge pairs in the higher $f_{DbCS}$ bins, peaking in the top 10$\%$ $f_{DbCS}$ bin. Within each $f_{DbCS}$ bin, the correlation is strongest for $n_5/s=0.2$ and progressively weaker for $n_5/s=0.1$ and 0.0, indicating a direct relationship between the CME signal injection and the $\gamma$-correlators. For the SS pairs, the $\gamma$-correlators are negative in the top $f_{DbCS}$ bins. The $\gamma$-correlator values are significantly higher in top $f_{DbCS}$ bins than those in the corresponding average centrality values, indicating strong CME signal in top $f_{DbCS}$ bins. The $\gamma$-correlator for both SS and OS pairs for the charge shuffle ($\gamma_{ChS}$) background increases significantly in the top $f_{DbCS}$ bins and are independent of the CME signal strength. The $\gamma$-correlators for the correlated background are consistent across all $f_{DbCS}$ bins for each AVFD set, with the highest values observed for $n_5/s=0.2$. The $\gamma$-correlators for $Ru+Ru$ and $Zr+Zr$ collisions show higher correlations compared to $Au+Au$ collisions as seen in Fig.~\ref{fig:gamma_fdbcs}, primarily due to increased background in the isobaric collisions. Notably, for the top 20$\%$ $f_{DbCS}$ bins, the magnitude of $\mid\gamma_{SS}\mid$ is approximately equal to $\mid\gamma_{OS}\mid$ across all data sets for $Ru+Ru$, $Zr+Zr$, and $Au+Au$ collisions, as well as their corresponding charge shuffle samples. This behavior is distinct from what is observed in Fig.~\ref{fig:gamma} for the overall centrality.\par
\begin{figure}
  \centering
  \includegraphics[width=.40\textwidth]{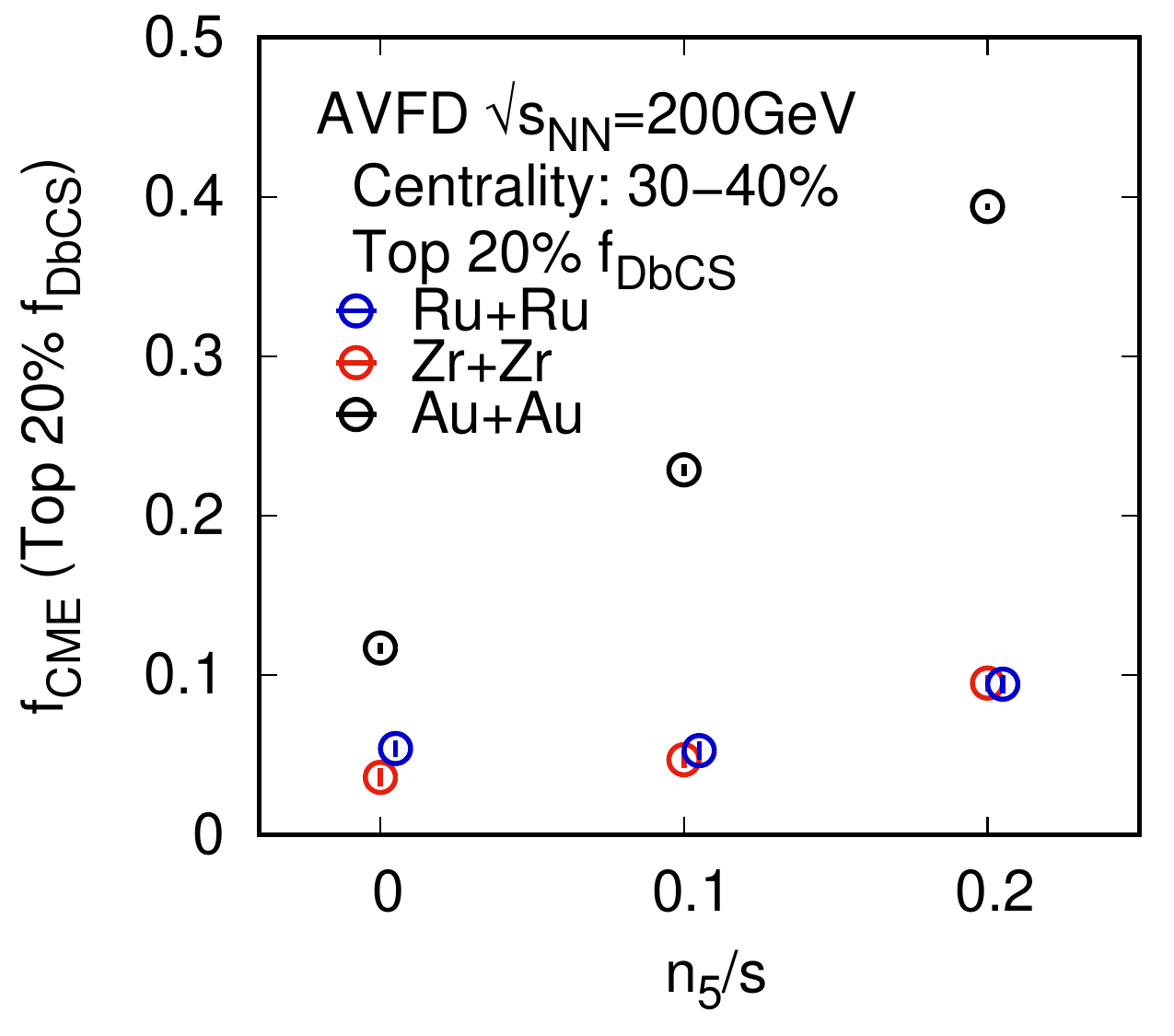}
  \caption{(Color online) $f_{CME}$ versus $n_5/s$ for $Ru+Ru$, $Zr+Zr$ and $Au+Au$ collisions at $\sqrt{s_{\mathrm NN}}$ = 200 GeV for top 20$\%$ $f_{DbCS}$ bins. Markers are slightly shifted along the x-axis for clarity. Statistical uncertainties are small and are within the marker size.}
  \label{fig:fcme}
\end{figure}
Two-particle $\delta$ correlators as a function of $f_{DbCS}$ for $Ru+Ru$ (top left), $Zr+Zr$ (top right), and $Au+Au$ (bottom) collisions for both opposite-sign and same-sign charged particles are presented in Fig.~\ref{fig:delta11}. Results indicate that $\delta_{OS}$ is negative while $\delta_{SS}$ is positive for the top 20$\%$ $f_{DbCS}$ bins. This is in contrast to the $\gamma$-correlators (Fig.~\ref{fig:gamma_fdbcs}), where $\gamma_{OS}$ is positive and $\gamma_{SS}$ is negative. These trends of $\gamma$ and $\delta$ correlators align with expectations for the CME-like events in top 20$\%$ $f_{DbCS}$ bins~\cite{Bzdak_2013}. Additionally, the $\delta$ correlators show a weak dependence on $n_5/s$.
\par
Figure~\ref{fig:InplanefDbCS} displays the in-plane and out-of-plane correlations for $Ru+Ru$ (top), $Zr+Zr$ (middle), and $Au+Au$ (bottom) collisions at $\sqrt{s_{\mathrm NN}}$ = 200 GeV for opposite-sign (left) and same-sign (right) charge pairs. The data reveals that opposite-sign correlations are stronger in the out-of-plane configuration, while in-plane correlations are weaker and show negative values in the top 20$\%$ $f_{DbCS}$ bins across all three collision types. Conversely, same-sign correlations are positive and demonstrate stronger out-of-plane correlations compared to in-plane correlations within the top 20$\%$ $f_{DbCS}$ bins. Furthermore, the out-of-plane correlations appear to increase with rising axial charge density per entropy.\par
The $\Delta\gamma$ (= $\gamma_{OS}$-$\gamma_{SS}$) plotted against different $f_{DbCS}$ percentile bins for $Ru+Ru$ (top left), $Zr+Zr$ (top right), and $Au+Au$ (bottom) collisions at $\sqrt{s_{\mathrm{NN}}}$ = 200 GeV are displayed in Fig.~\ref{fig:Deltagamma_fdbcs}. The comparisons of their corresponding charge shuffle ($\Delta\gamma_{ChS}$) and correlated backgrounds ($\Delta\gamma_{Corr}$) are also shown. As the CME signal increases, the $\Delta\gamma$ values rise across all $f_{DbCS}$ bins. The highest $\Delta\gamma$ values are found in the top 10$\%$ $f_{DbCS}$ bin. The $\Delta\gamma$ values for ChS backgrounds are nearly identical within statistical errors as expected for different $n_5/s$. The $\Delta\gamma$ values for the correlated backgrounds remain consistent across all $f_{DbCS}$ bins. The $\Delta\gamma$ is generally higher in isobaric collisions ($Ru+Ru$ and $Zr+Zr$) compared to $Au+Au$ collisions. This is attributed to the fact that $\Delta\gamma$ varies inversely with multiplicity~\cite{Lacey:2022plw}, which is higher in $Au+Au$ collisions, leading to lower $\Delta\gamma$ values. Additionally, it has been noted that $\Delta\gamma$ for the top 10$\%$ $f_{DbCS}$ bin is roughly ten times greater than values observed for the overall centrality (Fig.~\ref{fig:gamma} (bottom)) across all data sets including $n_5/s=0.0$ which contains only 33$\%$ LCC. This enhancement in top $f_{DbCS}$ bins is also reflected in the charge shuffle samples those represent background, although their values are nearly zero for overall centrality (Fig.~\ref{fig:gamma} (bottom)).\par
Based on the above observations concerning the three-particle correlators ($\gamma$), two-particle correlators ($\delta$), and in-/out-of-plane correlations, the top 20$\%$ of $f_{DbCS}$ events, which align with the expected CME signal~\cite{Bzdak_2013}, are identified as potential CME candidates. Consequently, the fraction of CME ($f_{CME}$) can be calculated using the following equation:
\begin{equation}
  \begin{split}
    f_{CME} & = 1 - \frac{\Delta\gamma_{Bkg}}{\Delta\gamma_{AVFD}} \\ 
    & \Delta\gamma_{Bkg} = \Delta\gamma_{ChS}+\Delta\gamma_{Corr}
  \end{split}
  \label{eq:CME}
\end{equation}

Figure~\ref{fig:fcme} shows the fraction of CME ($f_{CME}$) as a function of $n_5/s$ for $Au+Au$, $Ru+Ru$, and $Zr+Zr$ collisions. The results indicate that $f_{CME}$ increases with increasing externally injected CME signal. For $Au+Au$ collisions, $f_{CME}$ increases from 11.70$\pm$0.35$\%$ at $n_5/s=0.0$ to 39.30$\pm$0.20$\%$  at $n_5/s=0.2$, considering the top 20$\%$ $f_{DbCS}$ bins. In the $Ru+Ru$ collisions, $f_{CME}$ increases from 5.39$\pm$0.53$\%$ at $n_5/s=0.0$ to 9.43$\pm$0.56$\%$ at $n_5/s$=0.2. Similarly, for $Zr+Zr$ collisions, $f_{CME}$ rises from 3.58$\pm$0.59$\%$ at $n_5/s=0.0$ to 9.50$\pm$0.54$\%$ at $n_5/s=0.2$. Notably, even at $n_5/s=0.0$, $f_{CME}$ exhibits positive values. This suggests that the presence of 33$\%$ local charge conservation (LCC) in these samples mimics CME. For $n_5/s=0.2$, the $f_{CME}$ for $Au+Au$ collisions is 39.30$\pm$0.20$\%$ , which is roughly four times larger than the values for $Ru+Ru$ and $Zr+Zr$ collisions. \par
It is observed that in $Au+Au$ collisions, the $f_{CME}$ value doubles when $n_5/s$ increases from 0.0 to 0.1, and triples when $n_5/s$ reaches 0.2. However, in the case of isobaric collisions ($Ru+Ru$ and $Zr+Zr$), the increase in $f_{CME}$ is less pronounced. Notably, there is no increase in $f_{CME}$ for isobaric collisions when $n_5/s$ changes from 0.0 to 0.1. The small CME signal is difficult to distinguish in the presence of 33$\%$ LCC, as the background increases with decreasing multiplicity. Additionally, the results for $Ru+Ru$ and $Zr+Zr$ collisions are consistent within statistical errors. This consistency suggests that the increased magnetic field in $Ru+Ru$ collisions, compared to $Zr+Zr$, does not lead to a detectable increase in the CME signal, consistent with experimental observations~\cite{STAR_PRC105_2022}.
\section{Summary}
\label{sec:summary}
The AVFD model  generated $^{197}_{79}Au+^{197}_{79}Au$ and isobaric ($^{96}_{44}Ru+^{96}_{44}Ru$ and $^{96}_{40}Zr+^{96}_{40}Zr$) collisions with 33$\%$ LCC in each event for different CME signal strengths at $\sqrt{s_{NN}}=200$ GeV, have been extensively analyzed using the Sliding Dumbbell Method (SDM). The potential CME-like events identified through this method underwent scrutiny to ensure they exhibit the characteristics typical of CME events.
\par
In $Au+Au$ collisions, larger CME fraction ($f_{CME}$) is observed in sample with CME signal along with 33$\%$ LCC, and this fraction increases with increasing CME contribution. The $f_{CME}$ in $Au+Au$ collisions increases from 11.70$\pm$0.35$\%$ ($n_5/s=0.0$) to 39.30$\pm$0.20$\%$ ($n_5/s=0.2$) in the top 20$\%$ $f_{DbCS}$ bins.
It is worth noting that in top $10\%$ $f_{DbCS}$ bins, $\Delta\gamma$ increases significantly ($\sim$10 times) as compared to its values for overall centrality. This holds true even for the $n_5/s = 0$ sample, which represents the LCC background, as well as for the charge-shuffled sample. It is observed that the 33$\%$ LCC in a given sample mimics a CME-like signal.
However, in isobaric collisions, the increase in CME fraction with increasing CME signal is not observed for $n_5/s=0.1$, likely due to lower event multiplicities leading to significant background noise in these cases. Nevertheless, we do observe CME signal in samples with substantial injected CME ($n_5/s=0.2$).
However, no differences in the CME signals between the two isobars are noted, consistent with previous experimental findings. Therefore, the absence of an enhanced CME signal in $Ru+Ru$ collisions compared to $Zr+Zr$ collisions in experiments should not be interpreted as evidence that CME does not exist. This is clearly illustrated by the AVFD model, which incorporates the CME signal as well as 33$\%$ LCC and shows no increase in the CME signal for $Ru+Ru$ collisions over $Zr+Zr$ collisions.
\par
The SDM can be applied to experimental data on $Au+Au$, $Ru+Ru$, $Zr+Zr$, and $Pb+Pb$ collisions to validate the CME signal, as it enables the identification of potential CME-like candidates with a significantly higher CME fraction compared to the conventional approaches of searching within a fixed centrality range.
\begin{acknowledgments}
The authors sincerely thank Dr. Yufu Lin for generating and providing access to the AVFD events. The financial assistance from Department of Science $\&$ Technology, University Grants Commission, and Council of Scientific $\&$ Industrial Research, Government of India, is gratefully acknowledged. The authors are also thankful to the Panjab University and the Department of Physics for providing an academic environment and research facilities.
\end{acknowledgments}
\bibliographystyle{utphys}
\bibliography{biblio}

\end{document}